\documentclass[aps,floats,floatfix,showpacs,prd,superscriptaddress,onecolumn,nofootinbib]{revtex4}
\usepackage{graphicx, epsfig, amssymb} 
\usepackage{amsmath, amsfonts}
\usepackage{bm} 
\usepackage{color}
\usepackage{enumerate}
\usepackage[T1]{fontenc}
\usepackage{float}
\usepackage{array}
\usepackage{multirow}
\usepackage[multiple]{footmisc}

\makeatletter

\@ifundefined{textcolor}{}
{%
 \definecolor{BLACK}{gray}{0}
 \definecolor{WHITE}{gray}{1}
 \definecolor{RED}{rgb}{1,0,0}
 \definecolor{GREEN}{rgb}{0,1,0}
 \definecolor{BLUE}{rgb}{0,0,1}
 \definecolor{CYAN}{cmyk}{1,0,0,0}
 \definecolor{MAGENTA}{cmyk}{0,1,0,0}
 \definecolor{YELLOW}{cmyk}{0,0,1,0}
}

\makeatother

\newcommand{\be}{\begin{equation}}
\newcommand{\ee}{\end{equation}}                  
\newcommand{\bea}{\begin{eqnarray}}
\newcommand{\eea}{\end{eqnarray}}

\begin{document}

 
\title{AdS-Taub-NUT spacetimes and exact black bounces with scalar hair}

\author{José Barrientos}
\email{barrientos@math.cas.cz}
\affiliation{Institute of Mathematics of the Czech Academy of Sciences, \v{Z}itná 25, 11567 Praha 1, Czech Republic}
\affiliation{Departamento de Enseñanza de las Ciencias B\'asicas, Universidad Cat\'olica del Norte, Larrondo 1281, Coquimbo, Chile}

\author{Adolfo Cisterna}
\email{adolfo.cisterna.r@mail.pucv.cl}
\affiliation{Institute of Mathematics of the Czech Academy of Sciences, \v{Z}itná 25, 11567 Praha 1, Czech Republic}
\affiliation{Dipartimento di Fisica, Università di Trento,
Via Sommarive 14, 38123 Povo, Trentino, Italy.}
\affiliation{TIFPA - INFN, Via Sommarive 14, 38123 Povo, Trentino, Italy.}

\author{Nicolás Mora}
\email{nicolasmora@udec.cl}
\affiliation{Departamento de Física, Universidad de Concepción, Casilla 160-C, Concepción, Chile}

\author{Adriano Viganò}
\email{adriano.vigano@unimi.it}
\affiliation{Istituto Nazionale di Fisica Nucleare (INFN), Sezione di Milano
Via Celoria 16, I-20133 Milano, Italy}
\affiliation{Università degli Studi di Milano Via Celoria 16, I-20133 Milano, Italy}

\noindent 
\begin{abstract}

We present a new family of exact four-dimensional Taub-NUT spacetimes in Einstein-$\Lambda$ theory supplemented with a conformally coupled scalar field exhibiting a power-counting superrenormalizable potential. Our configurations are constructed in the following manner: A solution of a conformally coupled theory with a conformal potential, henceforth the seed $(g_{\mu\nu},\phi)$, is transformed by the action of a specific change of frame in addition with a simultaneous shift of the seed scalar field. The conformal factor of the transformation and the shift are both affine functions of the original scalar $\phi$. The new configuration, $(\bar{g}_{\mu\nu},\bar{\phi})$, solves the field equations of a conformally coupled theory with the extended aforementioned superrenormalizable potential, this under the presence of an effective cosmological constant. The new spectrum of solutions is notoriously enhanced with respect to the original seed containing regular black holes, wormholes, and bouncing cosmologies. We highlight the existence of two types of exact black bounces given by de Sitter and anti--de Sitter geometries that transit across three different configurations each. The de Sitter geometries transit from a regular black hole with event and cosmological horizons to a bouncing cosmology that connects two de Sitter Universes with different values of the asymptotic cosmological constant. An intermediate phase, which might be represented by two different configurations, takes place. These configurations are given by a de Sitter wormhole or by a bouncing cosmology that connects two de Sitter Universes, both under the presence of a cosmological horizon. On the other hand, the anti--de Sitter geometries transit from a regular black hole with inner and event horizons to a wormhole that connects two asymptotic boundaries with different constant curvatures. The intermediate phase is given in this case by an anti--de Sitter regular black hole with a single event horizon. This regular black hole might appear in two different configurations. As a regular anti--de Sitter black hole inside of an anti--de Sitter wormhole or as an anti--de Sitter regular black hole with a cosmological bounce in its interior. All these geometries are shown to be smoothly connected by the mass parameter only. Other standard stationary black holes, bouncing cosmologies and wormholes are also identified.

\end{abstract}


\maketitle


\section{Introduction}

The Taub-NUT solution of Einstein equations \cite{Taub:1950ez,Newman:1963yy} represents the simplest stationary generalization of the Schwarzschild black hole and it is well known for having intriguing topological features. Along with the standard mass parameter it also includes the so-called NUT charge, a continuous parameter usually regarded as a magnetic mass, basically the dual of the standard mass, that offers a natural realization of the gravitational counterpart of the electric-magnetic duality \cite{Dowker:1967zz}. The spacetime is thus interpreted, in direct analogy with the Dirac magnetic monopole, as the gravitational field of a dyon. The causal structure of the solution is wide, offering regions in which the spacetime is interpreted as a stationary black hole and regions in which is recognized as a nonsingular anisotropic cosmological model. These regions are respectively known as NUT and Taub. 
Due to the form of the nondiagonal components of the metric the spacetime is a priori pathological, specially in the NUT regions. It possesses a semi-infinite line singularity, usually called Misner string (the gravitational analogue of the Dirac string for the magnetic monopole) \cite{Misner:1963fr,Astefanesei:2004kn} and it contains regions plagued by closed timelike curves \cite{Griffiths:2009dfa,Hawking:1973uf}, both rendering the black hole interpretation misleading. 
Despite these undesirable features, this simple nonradiating spacetime deserves attention, a reason why different interpretations have emerged in order to study it. \\
\\
One interpretation, due to Misner \cite{Misner:1963fr}, regularizes the metric everywhere by introducing two coordinate patches that cover independently the north and south hemispheres of the symmetry axis and that smoothly connect at the equator. 
These patches are based on a change of the time coordinate that involves the periodic angular coordinate $\varphi$, i.e $t=t^{(\pm)}\mp C\varphi$, where $C$ is a constant proportional to the NUT charge that defines which semiaxis is being covered. 
Once an observer has access to the coordinates that cover the north hemisphere the Misner string takes place in the southern semiaxis and vice versa when the observer belongs to the patch covering the south hemisphere. The Misner string is then unseen.
In the overlap region both coordinates $t^{(\pm)}$ are related, and due to the periodicity of the azimuthal angle this ultimately implies that time is identified with a period proportional to the NUT charge\footnote{This approach is mathematically identical to the one that eliminates the Dirac string in the vector potential of the Dirac monopole. In there, the periodicity condition translates into the quantization of the electric charge.}. 
This results into the existence of closed timelike curves in the NUT regions. Moreover, under this periodicity condition the analytically extended Taub-NUT spacetime is either geodesically incomplete \cite{Misner:1963fr,Hawking:1973uf} or maximally extendable to a spacetime with no Haussdorff topology \cite{RGeroch,PHajicek}. Motivated by these considerations the Taub-NUT spacetime is usually studied in its Euclidean version where it represents a gravitational instanton, a gravitational cousin of Euclidean solutions of Yang-Mills theory \cite{Rajaraman,Shifman}. It is expected that these gravitational instantons play a similar role in theories of quantum gravity \cite{Esposito:1997wt} as the one played by Yang-Mills instantons in quantum field theory. Several models has been investigated in this direction \cite{Page:1978hdy,Iwai:1992zu,Witten:2009xu,Corral:2019leh,Arratia:2020hoy,Corral:2021xsu,Bueno:2018uoy}. \\
\\
On the other hand, there is Bonnor's interpretation, in which the Misner string is not avoided by any identification of the time coordinate and it is treated as a sort of topological defect usually regarded as a semi-infinite rod with spin \cite{BOnnor,SAck}. In fact, rewriting the Taub-NUT spacetime in Lewis-Weyl-Papapetrou coordinates it was pointed out in \cite{Stephani:2003tm} that NUT regions might represent the exterior field of a rotating source of which its angular momentum is measured by the NUT charge. 
The Misner string now represents an unremovable quasiregular singularity\footnote{Quasiregular in the sense that all curvature polynomials are still bounded.} and it naturally sows the questioning of the geodesic completeness of the spacetime. Moreover, regions containing closed timelike curves are still present.    
Nevertheless these disadvantages, it has been recently shown that the presence of the Misner string is not that catastrophic. In fact, in \cite{Clement:2015cxa,Clement:2015aka} the authors have shown that the topological defect of the Misner string is transparent to geodesic observers, and then there is no need to invoke any compactification of the time coordinate. In consequence, there are no obstructions for the analytic continuation of the spacetime through the horizons and then, due to the lack of curvature singularities, the whole Taub-NUT spacetime is geodesically complete. Moreover, it is also demonstrated that for certain values of $C$, basically the constant that locates the Misner string, geodesic observers do not encounter any causality violations, so the absence of closed timelike geodesics is ensured\footnote{For a detailed discussion about modern interpretations of the Taub-NUT spacetime we refer the readers to \cite{Tesis}.}.\\ 
\\
Among all gravitational solutions black holes represents the most iconic ones. Formed by gravitational collapse, one of the most astonishing features of black holes asserts that regardless the type of matter involved in the collapse they are finally characterized by a restricted set of quantities: mass, electromagnetic charges and angular momentum, namely, physical quantities satisfying a Gauss type law and in consequence measurable from infinity. Any other information about the original matter that formed the black hole is lost, either eaten up or expelled out from the black hole once equilibrium is reached. 
This statement, well-known as the no-hair conjecture \cite{Ruffini:1971bza}, is strongly supported in Einstein theory by the uniqueness of the Kerr-Newman family \cite{Carter:1971zc,Israel:1967wq,Israel:1967za,Wald:1971iw}. 
Hair is used to englobe all those properties of the collapsing matter that might depart from the mass, electromagnetic charges or angular momentum of the final configuration and that might represent a nontrivial trace of the type of matter involved in the gravitational collapse. 
The no-hair conjecture is a theory dependent statement and therefore it must be independently enunciated on all theories in which black holes are worth study. The simplest type of hair to be explored is the scalar one, basically complement Einstein theory with a minimally coupled scalar field that inherits the symmetries of the spacetime under scrutiny. An exact solution for this system was early known \cite{Janis:1969ivo}, although unphysical due to the presence of a naked singularity.
Bekenstein no-hair theorem \cite{Bekenstein:1995un,Bekenstein:1998aw} is the first mathematical realization of the no-hair conjecture and rules out the existence of black hole solutions with nontrivial scalar hair in Einstein theory with minimally coupled self-interacting scalar fields. Nonetheless, it also tangentially established the foundations to elude the absence of scalar hair by reaccommodating the hypothesis associated with the theorem.\\
\\
Along these lines the first black hole solution presented as a hairy scalar black hole, independently found by Bekenstein \cite{Bekenstein:1974sf} and by Bronnikov, Melnikov, and Bocharova \cite{BBM} (BBMB black hole), was proposed for Einstein theory supplemented with a conformally coupled scalar field. 
The solution possesses the causal structure of an extremal Reissner-Nordstr\"om black hole with a scalar field that diverges at the horizon, apparently with no dramatic consequences \cite{Bekenstein:1975ts}. 
It is usually introduced as a black hole in dimension four only \cite{Xanthopoulos:1992fm} and it relates to the one of \cite{Janis:1969ivo} by a conformal transformation, providing a precise example of how a change of frame substantially changes the geometric structure of a given solution\footnote{For a recent and detailed review of no-hair theorems and counterexamples to this conjecture we refer to \cite{Herdeiro:2015waa} and references therein.}. However, a deeper analysis \cite{Sudarsky:1997te} reveals that the divergence of the scalar field is actually harmful and that it is 
connected with an ill-defined stress tensor at the horizon. As a result the BBMB solution does not solves Einstein equations in an extended chart containing the event horizon and its interpretation as a genuine black hole is obscure. 
The situation, however, is relieved by the mere inclusion of a cosmological constant, which comes into play along with a conformal self-interacting potential for the scalar. A wider family of solutions is found for this case, the so-called Mart\'inez-Troncoso-Zanelli (MTZ) spacetimes \cite{Martinez:2002ru,Martinez:2005di}.
Here, the scalar field pole is hidden behind the event horizon and the solutions represent stealths and black holes with different topologies. Einstein equations are then everywhere solved in the domain of outer domains of communications, including the horizon and solving the undesired pathologies of the BBMB solution. 
Additionally, contrary to the BBMB black hole these solutions are not pathological in the minimally coupled frame. The integrability properties of these models have been well-studied since then, providing solutions for spacetimes with acceleration \cite{Charmousis:2009cm,Astorino:2013xxa}, NUT charge \cite{Bardoux:2013swa}, rotation \cite{Astorino:2014mda} and a full analysis of conformally coupled black holes in the entire Pleba\'nski-Demia\'nski family \cite{Anabalon:2009qt,Cisterna:2021xxq}. Moreover, several solutions have been found also in the presence of external magnetic fields \cite{Astorino:2013xc,Astorino:2013sfa}, in theories with higher order corrections \cite{Caceres:2020myr} and in metric-affine geometries \cite{Boudet:2020eyr}.\\
\\
In this context a particularly interesting solution is the one constructed in \cite{Anabalon:2012tu}. In there the usual four dimensional conformally coupled scalar theory is augmented with a self-interaction that deviates from the conformal potential by allowing the presence of a linear and cubic terms. The causal structure of the solution is notoriously enhanced with respect to the MTZ family \cite{Martinez:2002ru,Martinez:2005di} including regular black holes, wormholes and cosmological spacetimes with bounces\footnote{The vanishing cosmological constant case was early discussed in \cite{Barcelo:2000zf} and it represents a traversable wormhole.}. 
The inclusion of a Maxwell field was studied in \cite{Ayon-Beato:2015ada} where the authors have shown that the MTZ family and the solutions presented in \cite{Anabalon:2012tu} are related by a particular set of transformations; the simultaneous action of a conformal change of frame plus a specific shift of the initial scalar field configuration. Both, the conformal factor and the shifted scalar field are affine functions of the original scalar configuration. 
In consequence, the solutions contained in \cite{Anabalon:2012tu} are the transformed version of the MTZ solutions, this by means of the transformations provided in \cite{Ayon-Beato:2015ada}. The scheme, although different, works in analogy to the one that allows the construction of the BBMB black hole \cite{Bekenstein:1974sf} by applying a conformal transformation onto the originally pathological solution \cite{Janis:1969ivo}. In there, a solution of a minimally coupled scalar theory is translated into a solution of a conformally coupled theory by means of a conformal transformation. The set of transformations defined in \cite{Ayon-Beato:2015ada} transforms a solution of a conformally coupled theory into a solution of a new conformally coupled theory in which the scalar potential now deviates from the conformal potential by including all power-counting super-renormalizable contributions and where a new effective cosmological constant takes place. \\
\\
In the present work we face the construction of a new family of Taub-NUT spacetimes in the context of conformally coupled scalar-tensor theories. In a nutshell we construct the Taub-NUT version of the solutions described in \cite{Anabalon:2012tu,Ayon-Beato:2015ada}. Our starting point is the seed configuration studied in \cite{Bardoux:2013swa}, which by means of the transformations provided in \cite{Ayon-Beato:2015ada} is upgraded to represents a new set of geometries given by regular black holes, wormholes and bouncing cosmologies. Unnoticed in the previous works \cite{Anabalon:2012tu,Ayon-Beato:2015ada}, here we explicitly exploit the relevance of the new self-interaction: Under certain conditions it provides the existence of exact black bounce geometries that smoothly transit between regular black holes and wormholes and bouncing cosmologies. The transitions, as we shall observe, are controlled by the mass parameter only and do not require the presence of any new artificial scale into the theory. 
These type of solutions have recently attracted considerable attention, see \cite{Simpson:2018tsi,Simpson:2019oft} and references therein, most of them representing candidate spacetimes with interesting geometric properties but lacking from a well-defined action principle from which the equations of motion provide the desirable configurations. Here, we analyze from a heuristic point of view the geometric structure of these kind of exact spacetimes and we give a detailed analysis of every new backreaction, this accordingly with the sign of the cosmological constant and the different topologies of the base manifold. \\
\\
Our results organize as follows: Sec. II is devoted to provide preliminary results concerning with the construction of the solutions. First, we summarize the transformations presented in \cite{Ayon-Beato:2015ada} showing how a seed solution of a conformally coupled scalar theory is transformed into a solution of a conformally coupled scalar theory with an extended self-interacting potential. Then, we continue by giving a robust analysis of the seed solution \cite{Bardoux:2013swa}, this in order to explicitly determine how its original causal structures are enhanced by the action of the transformations.   
In Sec. III we present our novel solutions, discuss their main general properties and give a detailed analysis of every of these new geometries. We segment our analysis according to the sign of the cosmological constant and we describe every possible horizon geometry, namely, spherical, hyperbolic or planar. Several families of regular black holes, wormholes and bouncing cosmologies are described paying particular attention to those cases representing black bounces that transit across different geometries.  
We finally conclude in Sec. IV where we summarize our principal findings and where concrete extensions of the ideas proposed here are debated. The vanishing cosmological constant case is briefly described in the appendix.

\section{Preliminaries}

Before presenting our new solutions we start by providing a concise review of the technique developed in \cite{Ayon-Beato:2015ada} and the main features of the seed configuration \cite{Bardoux:2013swa}. As stated before, the authors have developed a technique by which it is possible to obtain new exact solutions in conformally coupled theories where the scalar self-interaction is no longer conformal invariant, this by simply applying a specific set of transformations onto a known solution, dubbed the seed, of a fully conformal invariant scalar theory. 
The transformations are nothing else than a change of frame, for which the conformal factor is an affine function of the scalar field seed configuration. A specific shift of the seed scalar field is simultaneously performed. The combined effect of both transformations results into an upgrade of the original action, the cosmological constant acquires an effective value while the scalar field self-interaction is enhanced (deviating from conformal invariance), containing for some specific spacetime dimensions all power-counting super-renormalizable contributions. The new solutions, which solve the field equations coming from the transformed action, contain a much richer causal structure as it has been already implicitly observed in \cite{Anabalon:2012tu}. The technique works in direct analogy with the usual conformal transformations that permit the integration of solutions in a standard conformally coupled scalar action starting from a minimally coupled scalar field theory \cite{Bekenstein:1974sf}. In what is next the transformations are specifically given and the transformed action is individualized. After the scheme behind the transformations is clear we proceed with the analysis of the Taub-NUT seed solution \cite{Bardoux:2013swa}, thus paving the road to understand how our novel solutions upgrade the geometric structure of the initial seed. 

\subsection{The transformation}

The starting point is the action of Einstein-$\Lambda$ theory supplemented by a self-interacting conformally coupled scalar field, 
\begin{equation}
S[g_{\mu\nu},\phi,A_\mu]=\int d^{D}x\sqrt{-g}\left[  \frac{R-2\Lambda}{2\kappa}-\frac{1}%
{2}\left(  \partial\phi\right)  ^{2}-\frac{\xi_D}{2}\phi^{2}R-V(\phi)-\frac{1}{4}F_{\mu\nu}F^{\mu\nu}\right].
\label{AP}%
\end{equation}
A Maxwell field is also considered. We define $\kappa=8\pi G$ and the nonminimal coupling with the curvature, as usual, is identified in arbitrary dimensions to the factor $\xi_D:=\frac{1}{4}\left(\frac{D-2}{D-1}\right)$. This value, up to a boundary term, dresses the scalar field action with invariance under the conformal transformations 
\begin{equation}
g_{\mu\nu}\rightarrow\Omega^2g_{\mu\nu}, \hspace{0.7cm} \phi\rightarrow\frac{\phi}{\Omega^{(D-2)/2}},
\end{equation}
with $\Omega$ a local function of the spacetime coordinates. The self-interaction is given by the conformal potential, which in arbitrary dimensions reads
\begin{equation}
V(\phi)=\alpha \phi^{\frac{2D}{D-2}}.
\end{equation}
Here $\alpha$ stands for a dimensionless coupling constant. The respective field equations, 
\begin{align}
G_{\mu\nu}+\Lambda g_{\mu\nu}&=\kappa\left[\partial_\mu\phi\partial_\nu\phi-\frac{1}{2}g_{\mu\nu}(\partial\phi)^2-\alpha g_{\mu\nu} \phi^{\frac{2D}{D-2}}+\xi_D(g_{\mu\nu}\Box-\nabla_\mu\nabla_\nu+G_{\mu\nu})\phi^2 \right] +\kappa\left(F_{\mu\lambda}F_{\nu}^{\lambda}-\frac{1}{4}g_{\mu\nu}F_{\lambda\rho}F^{\lambda\rho}\right)\nonumber\\
\Box\phi-\xi_DR\phi&=\alpha\frac{2D}{D-2}\phi^{\frac{D+2}{D-2}}\nonumber\\
\nabla_{\mu} F^{\mu\nu}&=0,   \label{seedEOM}
\end{align}
once solved, provide the corresponding seed solution $(g_{\mu\nu},\phi,A_\mu)$, which by means of the set of  transformations defined in \cite{Ayon-Beato:2015ada}
\begin{subequations}\label{eq:map}
\begin{align}
\bar{g}_{\mu\nu}&=\left(a\sqrt{\kappa\xi_{D}}\phi+1\right)^{\frac{4}{D-2}}g_{\mu\nu},
\label{eq:mapg}\\
\bar{\phi}&=\frac{1}{\sqrt{\kappa\xi_{D}}}\frac{\sqrt{\kappa\xi_{D}}\phi+a}
{a\sqrt{\kappa\xi_{D}}\phi+1},
\label{eq:mapPhi}\\
\bar{A}_{\mu}&=\sqrt{(1-a^2)}A_\mu \label{eq:mapA},
\end{align}
\end{subequations}
delivers a new conformally related solution $(\bar{g}_{\mu\nu},\bar{\phi},\bar{A_\mu})$ that solves the transformed field equations 
\begin{align}
\bar{G}_{\mu\nu}+\lambda \bar{g}_{\mu\nu}&=\kappa\left[\partial_\mu\bar{\phi}\partial_\nu\bar{\phi}-\frac{1}{2}\bar{g}_{\mu\nu}(\partial\bar{\phi})^2-\bar{g}_{\mu\nu}\bar{V}(\bar{\phi})+\xi_D(\bar{g}_{\mu\nu}\bar{\Box}-\bar{\nabla}_\mu\bar{\nabla}_\nu+\bar{G}_{\mu\nu})\bar{\phi}^2\right]+\kappa\left(\bar{F}_{\mu\lambda}\bar{F}_{\nu}^{\lambda}-\frac{1}{4}g_{\mu\nu}\bar{F}_{\lambda\rho}\bar{F}^{\lambda\rho}\right)\nonumber\\
\bar{\Box}\bar{\phi}-\xi_D\bar{R}\bar{\phi}&=\frac{\partial \bar{V}(\bar{\phi})}{\partial\bar{\phi}}\nonumber\\
\bar{\nabla}_\mu \bar{F}^{\mu\nu}&=0. \label{transEOM}
\end{align}
These equations of motion arise from the redefined action principle 
\begin{equation}
\bar{S}[\bar{g}_{\mu\nu},\bar{\phi},\bar{A}_\mu]=\int d^{D}x\sqrt{-\bar{g}}\left[  \frac{\bar{R}-2\lambda}{2\kappa}-\frac{1}%
{2}\left(  \partial\bar{\phi}\right)  ^{2}-\frac{\xi_D}{2}\bar{\phi}^{2}\bar{R}-\bar{V}(\bar{\phi})-\frac{1}{4}\bar{F}_{\mu\nu}\bar{F}^{\mu\nu}\right], 
\label{APP}%
\end{equation}
which is a consequence of transformations (\ref{eq:map}) acting on (\ref{AP}). 
The new action relates to the seed action by means of $\bar{S}[\bar{g},\bar{\phi},\bar{A}_\mu]=(1-a^2)S[g_{\mu\nu},\phi,A_\mu]$, maintaining unitarity as long as the shift parameter is constrained to $|a|<1$.
It must be noticed that the neat effect of the transformations on the seed action translates into the presence of a redefined cosmological constant 
\begin{align}
\lambda&=\frac{\kappa}{\left(1-a^{2}\right)^{\frac{D+2}{D-2}}}
\left[\frac{\Lambda}{\kappa}+
\alpha\left(-\frac{a}{\sqrt{\kappa\xi_{D}}}\right)^{\frac{2D}{D-2}}\right],\label{eq:barLambda}
\end{align}
along with an enhanced scalar potential defined by
\begin{align}
\bar{V}(\bar{\phi})&={}\frac{1}{\left(1-a^{2}\right)^{\frac{D+2}{D-2}}}
\Bigg\{\frac{\Lambda}{\kappa}
\left[\left(1-a\sqrt{\kappa\xi_{D}}\bar{\phi}\right)^{\frac{2D}{D-2}}-1\right]+\alpha\left[ \left(\bar{\phi}-\frac{a}{\sqrt{\kappa\xi_{D}}}\right)^{\frac{2D}{D-2}}
-\left(-\frac{a}{\sqrt{\kappa\xi_D}} \right)^{\frac{2D}{D-2}}\right]\Bigg\}\label{eq:potential}.
\end{align}
This new potential, which ceases to be conformally invariant, is superrenormalizable for integer values of the conformal power $2D/(D-2)$. 
This only occurs in three, four, and six dimensions. 
In addition, by construction this potential is defined such that no zeroth-order term appears, and then the new cosmological constant is precisely the one of (\ref{eq:barLambda}).
As we will shortly observe, every new coupling constant $\alpha_{2D/(D-2)}$ will be related to both, the original conformal coupling $\alpha$ and the corresponding seed cosmological constant $\Lambda$.  

\subsection{The seed configuration}  

We are interested on the study of new Taub-NUT solutions, in consequence we fix ourselves to dimension four where the seed solution for the conformally coupled theory (\ref{AP}) has been previously reported in \cite{Bardoux:2013swa}. In order to properly study the new solutions we shall present in the next section, we first provide a summary of the principal features of the seed metric, and in this way we pave the road to understand the causal structure of our novel solutions and their main differences with respect to solutions previously reported in the literature. 
The field equations (\ref{seedEOM}) are solved by the line element 
\begin{equation}
ds^2=g_{\mu\nu}dx^\mu dx^\nu=-F(r)(dt+\mathcal{B})^2+\frac{dr^2}{F(r)}+(r^2+n^2)d\Sigma_K^2,
\end{equation}
with 
\begin{equation}
F(r)=-\frac{\Lambda}{3}(r^2+n^2)+\frac{(K-\frac{4}{3}n^2\Lambda)(r-M)^2}{(r^2+n^2)}, \label{charmetric}
\end{equation}
and where each K\"ahler potential and the corresponding K\"ahler manifolds are given by 
\begin{equation}  \mathcal{B} =
\left\{ \begin{aligned} 
4n\sin^2\frac{\theta}{2} d\varphi \qquad &\text{when} \qquad d\Sigma^2_{(K=1)} = d \theta^2 + \sin^2\theta d \varphi^2  \\
n\theta^2 d\varphi \qquad &\text{when} \qquad d\Sigma^2_{(K=0)} = d \theta^2 + \theta^2 d \varphi^2 \\
4n\sinh^2\frac{\theta}{2} d\varphi \qquad &\text{when} \qquad d\Sigma^2_{(K=-1)} = d \theta^2 + \sinh^2\theta d \varphi^2.  \\
\end{aligned} \right. \label{solution-B}
\end{equation}
The configuration is completed by the following scalar and electromagnetic fields 
\begin{equation}
\phi(r)=\sqrt{-\frac{\Lambda}{6\alpha}}\frac{\sqrt{n^2+M^2}}{r-M}, \hspace{0.5cm} \mathcal{A}=\frac{Qr}{r^2+n^2}(dt+\mathcal{B}).   \label{charmfields}
\end{equation}
$Q$\footnote{The uncharged case will imply the tuning of the cosmological constant in terms of the coupling $\alpha$. In addition, the $\Lambda=\alpha=0$ case will represent the Taub-NUT extension of the BBMB black hole \cite{Bekenstein:1974sf,BBM}.} stands for the electric charge and two conditions must be met, $\Lambda\alpha<0$ and 
\begin{equation}
Q^2=\frac{1}{18}\frac{(n^2+M^2)(K-\frac{4}{3}n^2\Lambda)(\kappa\Lambda+36\alpha)}{\kappa\alpha}.
\end{equation}
Reality of the electric charge finally requires  
\begin{align}
&(K-\frac{4}{3}n^2\Lambda)>0 \hspace{0.2cm} \text{and} \hspace{0.2cm} \frac{(\kappa\Lambda+36\alpha)}{\kappa\alpha}>0 \label{con1}, \hspace{0.1cm} or \\ 
&(K-\frac{4}{3}n^2\Lambda)<0 \hspace{0.2cm} \text{and} \hspace{0.2cm} \frac{(\kappa\Lambda+36\alpha)}{\kappa\alpha}<0\label{con2}.
\end{align}
Notice that we have applied the coordinate transformation $t\rightarrow t+2n\varphi$ onto the original solution presented in \cite{Bardoux:2013swa}. This allows to remove one of the singular semiaxis and localize the Misner string at the southern semiaxis $(\theta=\pi)$ only. Moreover, these coordinates provides a simpler setup in which to study the appearance of generic closed timelike curves associated with the change of sign of the metric component $g_{\varphi\varphi}$. 
From hereon we assume a combination of $\alpha$ and $\Lambda$ such that the gauge and scalar fields remain real. \\
 \\
The spacetime inherits the asymptotic behavior of Taub-NUT spaces, namely, with or without a cosmological constant the spacetime is not globally asymptotically (A)dS neither globally asymptotically flat. This, as it is known, is due to the presence of the nondiagonal components of the metric, basically the presence of the NUT parameter that turns on the Riemann tensor components associated with the magnetic part of the Weyl tensor. Nevertheless, the electric part of the Weyl tensor provides the associated Riemann tensor components, $\bar{R}^{\mu\nu}_{\,\,\sigma\rho}$, with the usual asymptotic $\lim_{r\rightarrow\infty}\bar{R}^{\mu\nu}_{\,\,\sigma\rho}=\frac{\Lambda}{3}\delta^{\mu\nu}_{\,\,\sigma\rho}$. 
The spacetime is free of curvature singularities, however the Misner string defect persists for a spherical foliation. Notwithstanding this, leaning on Bonnor's interpretation \cite{BOnnor,SAck} and the recent works \cite{Clement:2015cxa,Clement:2015aka} the spacetime does not exhibits any pathologies for free falling observers whose backreaction is negligible. 
The spacetime is taken as geodesically complete, and in consequence the radial coordinate takes values on the full range from $-\infty$ to $+\infty$. The mass parameter is then positive defined, negative values are englobed in the change of coordinate $r\rightarrow-r$.  
Despite the presence of the Misner string we analyze the causal structure of the seed metric for each of the horizon geometries. The metric function (\ref{charmetric}) possesses four roots 
\begin{align}
r_{--}&=\frac{\bar{l}}2\left( -1-\sqrt{1-4\frac{n^2}{\bar{l}^2}+4\frac{M}{\bar{l}}} \right)\\
r_{-}&=\frac{\bar{l}}2\left( -1+\sqrt{1-4\frac{n^2}{\bar{l}^2}+4\frac{M}{\bar{l}}} \right) \\
r_+&=\frac{\bar{l}}2\left( 1-\sqrt{1-4\frac{n^2}{\bar{l}^2}-4\frac{M}{\bar{l}}} \right)\\
r_{++}&= \frac{\bar{l}}2\left( 1+\sqrt{1-4\frac{n^2}{\bar{l}^2}-4\frac{M}{\bar{l}}} \right),  \label{horizons}
\end{align}
which might define up to four Killing horizons. Here, accordingly with the sign of the cosmological constant we have made use of the following conventions
\begin{align}
\Lambda>0& \rightarrow \bar{l}=\sqrt{3K/\Lambda-4n^2} \\
\Lambda<0& \rightarrow \bar{l}=\sqrt{-(-3K/\Lambda+4n^2)}.  \label{Lcon}
\end{align}
The location of these horizons satisfies $r_{--}<0<r_{-}<M<r_{+}<r_{++}$, where $r=M$ defines the point in which the scalar field diverges. The presence of the scalar field pole is a common feature in solutions of conformally coupled theories of the kind considered here for scalar fields that are radially dependent only. 
All roots of the lapse function $F(r)$ remain real as long as the mass parameter obeys $0<M<\frac{\bar{l}}{4}-\frac{n^2}{\bar{l}}$, reality condition that is subjected to the reality of $\bar{l}$. In addition, the mass parameter is taken to be always positive and then $\frac{\bar{l}}{4}-\frac{n^2}{\bar{l}}>0$ must be met. This case is the most generic one and implies the existence of four Killing horizons. 
One direct difference with respect to the static counterpart is the presence of a solution with a planar horizon, which in the absence of the NUT parameter requires the presence of extra matter fields \cite{Bardoux:2012aw}. For each of the possible horizon geometries the following features are sketched out:\\
\begin{enumerate}[i)]
\item $K=1$, spherical foliation:\\
\vspace{0.2 cm}\\
$\Lambda>0$: The reality of $\bar{l}$ and the condition $\frac{\bar{l}}{4}-\frac{n^2}{\bar{l}}>0$ are simultaneously achieved by considering $\Lambda<\frac{3}{8n^2}$. This allows the generic existence of Killing horizons. In addition, $\alpha<0$ provides a real scalar field configuration.
Thus, for $0<M<\frac{\bar{l}}{4}-\frac{n^2}{\bar{l}}$ there are four Killing horizons. Both, $r_{--}$ and $r_{++}$, correspond to cosmological horizons beyond which the metric becomes time dependent. The scalar field pole is hidden behind the event horizon $r_{+}$. The region between $r_{+}$ and $r_{++}$ is recognized as the exterior of a regular stationary black hole, however under the presence of a pathological semiaxis of rotation at $\theta=\pi$ and due to the generic existence of closed timelike curves its interpretation is questionable. Notwithstanding this, the existence of closed timelike curves is more subtle under the presence of the conformally coupled scalar field. These might be avoided by restricting the parameter space determined by the cosmological constant and the NUT charge, this at the price of changing the interpretation of the spacetime metric. It can be seen as follows: the $g_{\varphi\varphi}$ metric component is given by 
\begin{equation}
g_{\varphi\varphi}=\left[(r^2+n^2)\sin^2\theta-16n^2F(r)\sin^4\frac{\theta}{2}\right].
\end{equation}
Then, in order to maintain the Riemannian nature of the two dimensional metric given by $t=r=constant$ it is required to hold
\begin{equation}
\frac{4n^2F(r)}{r^2+n^2}<\ -1. 
\end{equation}
After some algebraic manipulations it is possible to show that\footnote{This is an exclusive property of the conformally coupled scalar case we consider here. It is mostly due to the quasiextremal form that the scalar field provides to the metric function.} 
\begin{equation}
\left(1-\frac{4}{3}n^2\Lambda\right)\left[1+4n^2\left(\frac{r-M}{r^2+n^2}\right)^2\right]< 0,
\end{equation}
providing the complete absence of closed timelike curves as soon as $\Lambda>3/4n^2$. This condition does not overlap with the requirement of a real $\bar{l}$ and in consequence under this condition the metric function is devoid of Killing horizons, all roots of $F(r)$ become complex. 
Therefore, the metric function $F(r)$ is everywhere negative, reason why the metric component $g_{\varphi\varphi}$ remains always positive. Then, the spacetime represents a cosmological model in which the scalar field pole is uncovered. \\
It is instructive to analyze the case in which the mass parameter satisfies $M>\frac{\bar{l}}{4}-\frac{n^2}{\bar{l}}$. 
In this case the roots $r_+$ and $r_{++}$ become complex while $r_-$ and $r_{--}$ remain real. It is not possible to make disappear all Killing horizons at the same time. Moreover, the scalar field pole stays uncovered. The metric function $F(r)$ is everywhere negative outside $r_-$ and the spacetime there, although pathological, possesses a cosmological behavior. 
This will be particularly interesting when presenting our novel solutions. We will see that after applying the transformations previously explained this spacetime will become a bouncing cosmology with no pole for the scalar field and with no pathologies associated to the presence of closed timelike curves. In addition, under specific circumstances a black bounce transition in between a regular black hole and a bouncing cosmology will be observable. \\
It is important to stress that in spite of this the Misner string is still present and that ultimately this obstruction, as shown in \cite{Clement:2015cxa,Clement:2015aka}, can be circumvented by considering geodesic observers only.
More appealing for spherical foliation is the Euclidean case, for which the Misner string can be completely removed by the Misner approach and for which the presence of closed timelike curves is not a pathology. This case will be further analyzed in \cite{eu-case}. 
\vspace{0.2 cm}\\
$\Lambda<0$: For a negative cosmological constant relation (\ref{Lcon}) implies always a complex $\bar{l}$, and in consequence the complete absence of Killing horizons no matter the values the mass parameter can take. The scalar field is real for $\alpha>0$ and its pole remains uncovered. The metric function $F(r)$ is always positive and the radial coordinate remains spacelike.
Both, a singular semiaxis and closed timelike curves are present.\\
\vspace{0.2 cm}\\
\item $K= -1$, hyperbolic foliation:\\ 
\vspace{0.2 cm}\\
$\Lambda>0$: In this case a positive cosmological constant implies the absence of Killing horizons, no real roots for $F(r)$. The metric function is everywhere negative, and in consequence the spacetime inherits a cosmological nature. The scalar field is real for $\alpha<0$ and its pole is uncovered. The hyperbolic geometry, in contrast to the previous spherical case, does not deal with the presence of the singular semiaxes, namely, Misner strings are not present \cite{Astefanesei:2004kn}. Then, only the presence of closed timelike curves must be analyzed. The azimuthal angle component of the metric now reads 
\begin{equation}
g_{\varphi\varphi}=\left[(r^2+n^2)\sinh^2\theta-16n^2F(r)\sinh^4\frac{\theta}{2}\right].
\end{equation}
Following a similar computation to the one explained for the spherical case it is shown that the absence of closed timelike curves is ensured by satisfying 
\begin{equation}
-\left(1+\frac{4}{3}n^2\Lambda\right)\left[1+4n^2\left(\frac{r-M}{r^2+n^2}\right)^2\right]< 0, \label{CTChyp}
\end{equation}
condition always achieved by positive cosmological constant. \\
\vspace{0.2 cm}\\
$\Lambda<0$: For a negative cosmological constant the reality of $\bar{l}$ and the positivity of the combination $\frac{\bar{l}}{4}-\frac{n^2}{\bar{l}}$ are ensured by $|\Lambda|<\frac{3}{8n^2}$, where we have considered $\Lambda=-|\Lambda|$. $\alpha$ is taken to be positive. Hence, for $0<M<\frac{\bar{l}}{4}-\frac{n^2}{\bar{l}}$ four Killing horizons take place, none of them of cosmological nature. The spacetime between the event horizon $r_{++}$ and asymptotic infinity is identified as the exterior of a regular stationary topological black hole where no Misner string takes place. Moreover, this case offers the complete absence of closed timelike curves as soon as $|\Lambda|<3/4n^2$, condition that overlaps with the generic existence of Killing horizons. This spacetime represents a well-behaved stationary black hole, free of Misner strings as well as free of causality violations for any observer.
On the other hand, for $M>\frac{\bar{l}}{4}-\frac{n^2}{\bar{l}}$ both roots $r_+$ and $r_{++}$ become complex and the metric function $F(r)$ is everywhere positive outside $r_-$. The scalar field pole is uncovered. In the next section we will observe how this geometry transforms into a wormhole spacetime with no associated pathologies, namely, Misner strings nor closed timelike curves, and how under specific conditions provides a black bounce that smoothly connects a regular black hole with an anti--de Sitter wormhole. \\
\vspace{0.2 cm}\\
\item $K= 0$, planar foliation:\\ 
\vspace{0.2 cm}\\
This case is particular, despite the value of the cosmological constant $F(r)$ is always devoid of real roots. Then, the spacetime is either radially dependent or time dependent according to $\Lambda<0$ or $\Lambda>0$, respectively. Accordingly, $\alpha>0$ or $\alpha<0$ and in both cases the scalar field pole is uncovered. Misner strings are pulled out to infinity, however closed timelike curves generically occur. 
\end{enumerate}

The case of vanishing cosmological constant delivers, for $K=1$, the well-known Taub-NUT extension of the BBMB black hole. Its structure and further analysis can be extracted from \cite{Bardoux:2013swa}. We will briefly comment on this case in the appendix. In there, we will explicitly show how it will represent the charged Taub-NUT extension of Barcelo's wormhole \cite{Barcelo:2000zf}.\\
\\
It must be noticed that the presence of Misner strings and of the causal pathologies associated with closed timelike curves can be generically circumvented only when the horizon is locally isomorphic to $\mathcal{H}^2$, as long as condition (\ref{CTChyp}) is met. This behavior is due to the presence of the conformally coupled scalar field and the extremal form it provides to the metric. However, notice that according to Bonnor's interpretation of the Misner string the spherical and planar solutions are not \emph{a priori}  pathological for geodesic observers. As it was previously explained, by a simple change of coordinates and a proper constraint on the constant $C$ \cite{Clement:2015cxa,Clement:2015aka}, the Misner string becomes transparent for geodesic observers and even more no closed timelike geodesics take place. This open a window in which all cases, no matter the presence of these pathologies, might be interpreted as black holes when it corresponds. Accordingly, the spherical and planar cases are worth study. We follow this interpretation and in consequence we will provide a detailed description of all cases, $K=\pm1,0$.

\section{solutions}

Once the main properties of the seed solution (\ref{charmetric}) are known it becomes straightforward to understand how the new solutions, the ones obtained by applying the set of transformations (\ref{eq:map}), upgrade the seed spacetime to a wider family of geometries. Obtained by applying the map (\ref{eq:map}) into the seed configuration (\ref{charmetric}) and (\ref{charmfields}) and solving in consequence the field equations coming from (\ref{APP}), our new spacetime takes the form
\begin{equation}
\begin{aligned}
d\bar{s}^2&=\frac{(ab\sqrt{n^2+M^2}+r-M)^2}{(r-M)^2}\left[-\left(-\frac{\Lambda}{3}(r^2+n^2)+\frac{(K-\frac{4}{3}n^2\Lambda)(r-M)^2}{(r^2+n^2)}\right)(dt+\mathcal{B})^2\right.\\
&\qquad\left.+\frac{dr^2}{\left(-\frac{\Lambda}{3}(r^2+n^2)+\frac{(K-\frac{4}{3}n^2\Lambda)(r-M)^2}{(r^2+n^2)}\right)}+(r^2+n^2)d\Sigma_{K}^2\right], \hspace{0.4cm} \bar{A}=\frac{\bar{Q}r}{r^2+n^2}(dt+\mathcal{B}),\\
\\
\bar{\phi}(r)&=\left(\frac{6}{\kappa}\right)^{1/2}\frac{b\sqrt{n^2+M^2}+a(r-M)}{ab\sqrt{n^2+M^2}+r-M},\hspace{0.8cm} b=\sqrt{\frac{-\kappa\Lambda}{36\alpha}},\hspace{0.8cm}\lambda=\frac{\kappa\Lambda+36\alpha a^4}{\kappa(1-a^2)^3}. 
\end{aligned} \label{fullsol}
\end{equation}
Notice that the condition $\alpha\Lambda<0$ ensures the reality of the parameter $b$.
In four dimensions the conformal power is an integer and in consequence the scalar potential (\ref{eq:potential}) contains all power-counting superrenormalizable contributions
\begin{equation}
\bar{V}(\bar{\phi})=\alpha_1\bar{\phi}+\alpha_2\bar{\phi}^2+\alpha_3\bar{\phi}^3+\alpha_4\bar{\phi}^4, 
\end{equation}
where consistence with all field equations requires to fix the coupling constants as 
\begin{align}
\alpha_1 = -\frac{2\sqrt{6}}{3}\frac{a(\Lambda\kappa+36a^2\alpha)}{\kappa^{\frac{3}{2}}(1-a^2)^3}, \hspace{0.1cm}
\alpha_2 = \frac{a^2(\Lambda\kappa+36\alpha)}{\kappa(1-a^2)^3} , \hspace{0.1cm}
\alpha_3 = -\frac{\sqrt{6}}{9}\frac{a(\Lambda a^2\kappa+36\alpha)}{\sqrt{\kappa}(1-a^2)^3} , \hspace{0.1cm}
\alpha_4 = \frac{1}{36}\frac{(a^4\kappa\Lambda+36\alpha)}{(1-a^2)^3}.
\end{align}
Notice that, beyond the superrenormalizable nature of the potential, each of the contributions make it worth studying. The conformal potential is itself interesting due to its Weyl invariance while on the other hand the quadratic contribution represents a mass term. More unconventional are the presence of the linear and cubic terms. Nonetheless, realistic models of a scalar field theory, as it is the case of the neutral pion $\pi_0$ and its sigma model description, do include these type of potentials, see for instance \cite{Donoghue:1992dd}.
Finally and in analogy with the seed configuration, the solution is closed by imposing the relation 
\begin{equation}
\bar{Q}^2 = \frac{1}{18}\frac{(n^2+M^2)(1-a^2)(K-\frac{4n^2\Lambda}{3})(\Lambda\kappa+36\alpha)}{\alpha\kappa}, \label{conQ}
\end{equation}
which in agreement with transformation (\ref{eq:mapA}) is nothing else than consider $\bar{Q}^2=(1-a^2)Q^2$. In consequence $\bar{Q}$ remains real as long as any of the constraints (\ref{con1}) and (\ref{con2}) is respected. \\
\\
The analysis of these new geometries starts by noticing two important differences with respect to the seed solution. These are the appearance of a curvature singularity and the emergence of new asymptotic regions, due to the pole of the conformal factor, that might split the spacetime into two causally disconnected different regions. 
In contrast with the seed metric and despite the presence of the NUT charge, which usually softens the geometry of Taub-NUT spacetimes, a curvature singularity located at 
\begin{equation}
r_0=M-ab\sqrt{n^2+M^2},  \label{singu} 
\end{equation}
emerges. This singularity arises due to the presence of the conformal factor and it precisely corresponds to the point in which the latter vanishes. 
Thus, beyond the interpretation of the Misner string the spacetime ends at the singularity $r_0$ which might be located either in the positive or negative range of the radial coordinate, this depending on the signs of $M$, $a$ and $b$. For the curvature singularity to be located at positives values of $r$ we observe, 
\begin{equation}  r_0^{+} >0 \rightarrow
\left\{ \begin{aligned} 
M>\frac{abn}{\sqrt{1-a^2b^2}}\qquad &\text{and}\qquad 0<ab<1\\
M>0 \qquad &\text{and} \qquad ab<0 \\
0<|M|<\frac{|ab|n}{\sqrt{1-|ab|^2}}  \qquad &\text{and} \qquad 0<|ab|<1\qquad \text{where}\qquad M=-|M|, ab=-|ab|.  \\
\end{aligned} \right. \label{poscuv}
\end{equation}
On the other hand, the singularity takes place at negative values of the radial coordinate if, 
\begin{equation}  r_0^{-} <0 \rightarrow
\left\{ \begin{aligned} 
0<M<\frac{abn}{\sqrt{1-a^2b^2}}\qquad &\text{and}\qquad 0<ab<1\\
|M|>\frac{|ab|n}{\sqrt{1-|ab|^2}}\qquad &\text{and}\qquad 0<|ab|<1\qquad \text{where}\qquad M=-|M|, ab=-|ab|.\\
|M|>0 \qquad &\text{and} \qquad ab>0.  \\
\end{aligned} \right.  \label{negcuv}
\end{equation}
Thus, the range of the radial coordinate is given by $(r_0^{\pm},\infty)$, where we have dubbed $r_0^{\pm}$ the positive and negative possible locations of the singularity. \\
\\
While the curvature singularity defines the domain of the radial coordinate, and basically the extension of the spacetime, the conformal factor defines the structure of the asymptotic regions. The presence of the asymptotic region at plus infinity is clear, nevertheless as already mentioned the conformal factor may exhibits a pole which eventually would include a new asymptotic region that drastically changes the causal structure of the solutions. It is then necessary to analyze the position of the conformal factor pole for both cases, whenever $r_0^+$ or $r_0^-$ takes place. Then, it is mandatory to understand how the singularity, the eventual new asymptotic region, and the Killing horizons are located with respect to each other. \\
\\
According to these observations, and before giving a detailed analysis of every of the geometries contained in (\ref{fullsol}), a few comments are in order: 
\begin{enumerate}[i)]
\item Although the new metric $\bar{g}_{\mu\nu}$ is defined through a change of frame of the seed metric $g_{\mu\nu}$ and in consequence inherits most of its features, the conformal factor dramatically changes the causal structure of the spacetime. 
As already mentioned its root introduces a curvature singularity (\ref{singu}), while on the other hand may drastically change the causal structure of the spacetime including a new asymptotic region at the point in which it diverges. The position of this pole depends on the sign of the mass parameter and accordingly to the location of the curvature singularity provides the following spacetimes:
\begin{itemize}
\item[--] $r_0^+$: The domain of the radial coordinate is given by $(r_{0}^{+},\infty)$. On the other hand, the conformal factor exhibits a pole either at $r=M$ or $r=-|M|$ depending if the mass parameter is positive or negative. For $M>0$ and, as long as $r_0^+<r=M$, a new asymptotic region emerges dividing the spacetime into two regions we have dubbed $\mathcal{R}^{-}\in(r^{+}_0,M)$ and $\mathcal{R}^{+}\in(M,\infty)$. For observers located in $\mathcal{R}^{+}$ the access to region $\mathcal{R}^{-}$ is not allowed, being this portion of the spacetime replaced by a new asymptotic region with its own constant curvature. Similarly occurs for observers located in $\mathcal{R}^{-}$, their access to $\mathcal{R}^{+}$ is not permited. This happens for the first of the conditions in (\ref{poscuv}). For the second case contained in (\ref{poscuv}) we observe that $r_0^+>r=M$, and in consequence the pole of the conformal factor does not belong to the domain of the radial coordinate. Similarly occurs for $M<0$, then the pole is located at $r=-|M|$ and again does not belong to the possible values the radial coordinate can take. Thus, for the last two cases exposed in (\ref{poscuv}) no extra asymptotic region takes place and the region $(r_{0}^{+},\infty)$ is not divided. 
\vspace{0.2cm}
\item[--] $r_0^-$: The domain of the radial coordinate is given by $(r_{0}^{-},\infty)$. Again, the pole of the conformal factor depends on the sign of the mass parameter, given either by $r=M$ or $r=-|M|$. For the first of the conditions contained in (\ref{negcuv}) the mass parameter is positive and in consequence the pole located at $r=M$ satisfies $r=M>r_0^-$. This pole induces the presence of a new asymptotic region which divides the spacetime into the regions $\mathcal{R}^{-}\in(r^{-}_0,M)$ and $\mathcal{R}^{+}\in(M,\infty)$, which as in the previous case are causally disconnected. For the second case of (\ref{negcuv}) we observe that $r_0^->r=-|M|$ and then the pole of the conformal factor does not belong to the domain $(r_{0}^{-},\infty)$. The spacetime is not divided in this case. Finally, for the third condition we observe that the singularity satisfies $r_0^-<r=-|M|$, and then the pole of the conformal factor induces a new asymptotic region, this time located at $r=-|M|$, that splits the spacetime into the regions $\mathcal{R}^{-}\in(r^{-}_0,-|M|)$ and $\mathcal{R}^{+}\in(-|M|,\infty)$.
\vspace{0.2cm}
\item[--] These spacetimes might have different geometric interpretations depending on the space of parameters under consideration and on the relative position of the Killing horizons that this set of parameters defines. As matter of fact, the position of the Killing horizons is affected by the sign of the mass parameter, and in consequence their distribution varies from case to case accordingly to the position of the singularity and the position of an eventual new asymptotic region. 
\end{itemize}

\item For the cases in which the conformal factor induces a splitting of the spacetime we observe that the new asymptotic regions possess their own constant curvature with respect to asymptotic infinity. As inherited from the seed solution, the asymptotic behavior of the metric is of the Taub-NUT type and then it is only locally asymptotically equivalent to (anti)-de Sitter spacetime. The Riemann tensor components associated to the electric part of the Weyl tensor behave as 
\begin{align}
\lim_{\stackrel{r\rightarrow\infty}{M>0}}\bar{R}^{\mu\nu}_{\,\,\sigma\rho}&=\lim_{\stackrel{r\rightarrow\infty}{M<0}}\bar{R}^{\mu\nu}_{\,\,\sigma\rho}=\frac{\Lambda}{3}\delta^{\mu\nu}_{\,\,\sigma\rho},\nonumber\\
\lim_{r\rightarrow M>0}\bar{R}^{\mu\nu}_{\,\,\sigma\rho}&=\lim_{r\rightarrow M<0}\bar{R}^{\mu\nu}_{\,\,\sigma\rho}=\frac{\Lambda}{3a^2b^2}\delta^{\mu\nu}_{\,\,\sigma\rho}. \label{ctecuv}
\end{align}


\item The scalar field configuration diverges precisely at the curvature singularity, namely, at the zero of the conformal factor. Then, the scalar field is unbounded either at $r_0^{-}$ or $r_0^{+}$, depending on the case under consideration. The expectation value of the scalar on each of the possible boundaries, for positive and negative mass parameter, is given by $\bar{\phi}(M)\sim\sqrt{\frac{6}{\kappa}}\frac{1}{a}$ and $\bar{\phi}(\infty)\sim\sqrt{\frac{6}{\kappa}}a$.

\item The Killing horizons of these geometries are still given by (\ref{horizons}), namely, the Killing horizons of the seed metric. However, their relative locations as well as their location with respect to the new asymptotic region and the curvature singularities depend now on the sign of the mass parameter.  

\item The $a\rightarrow0$ limit, of course, brings the solutions into the seed \cite{Bardoux:2013swa}. On the other hand, keeping $a$ but going to the vanishing NUT charge case, the families of solutions described in \cite{Anabalon:2012tu} and  \cite{Ayon-Beato:2015ada} are recovered depending on whether or not the electric charge is present. In the vanishing $\Lambda$ case the latter two solutions corresponds to Barcelo's solution \cite{Barcelo:2000zf} with or without charge. Finally, for $n=a=0$, the BBMB \cite{Bekenstein:1974sf,BBM} and MTZ  \cite{Martinez:2002ru,Martinez:2005di} solutions are recovered depending on the absence or the presence of a cosmological constant and the corresponding self-interaction. 
\end{enumerate}
Some of these general considerations are holistically shared by the $n\rightarrow 0$ solutions \cite{Anabalon:2012tu} and  \cite{Ayon-Beato:2015ada}. We refer to \cite{Anabalon:2012tu} for more details regarding generic features of these kind of geometries in a more simple setup. 
In the next subsections we proceed with the analysis of the causal structures contained in our new solution (\ref{fullsol}), for positive and negative cosmological constant and for spherical, hyperbolic and planar base manifolds. 

\subsection{$\Lambda\neq0$ solutions}

In agreement with the analysis performed on the seed configuration, here we connect with a specific analysis of each of the new geometries pertaining to (\ref{fullsol}). The analysis is carried out in the following fashion: First it is defined the sign of the seed cosmological constant $\Lambda$ and the topology of the black hole horizon $K$. Then, for every $(\Lambda, K)$ combination we  separately describe the cases with curvature singularities at $r_{0}^{+}$ and $r_{0}^{-}$. We explore all the possibilities contained in (\ref{poscuv}) and (\ref{negcuv}). 
Due to the presence of these curvature singularities the mass parameter takes positive and negatives values. This implies that the appearance of all Killing horizons is more involved. As stated for the seed solution it is needed to ensure the reality of the quantity $\bar{l}$, then the generic existence of these horizons will depend on the values the mass parameter can take. For positive and negative values of $M$ we have: \\
\\
$M>0$: In this case it is observed that all four Killing horizons appear as long as the mass parameter satisfies $0<M<\frac{\bar{l}}{4}-\frac{n^2}{\bar{l}}$. This condition must be complemented with two requirements: $\bar{l}\in\mathbb{R}$ and $\frac{\bar{l}}{4}-\frac{n^2}{\bar{l}}>0$. Both requirements are fulfilled by 
$0<|\Lambda|<3/8n^2$ for positive and negative cosmological constant. \\
\\
$M<0$: In this case the situation is more subtle. Not all the roots for $F(r)$ are real simultaneously, thus not all Killing horizons appear at the same time. It is observed that both horizons $r_+$ and $r_{++}$ exist for $|M|>\frac{n^2}{\bar{l}}-\frac{\bar{l}}{4}$. Then conditions $\bar{l}\in\mathbb{R}$ and $\frac{n^2}{\bar{l}}-\frac{\bar{l}}{4}>0$ are simultaneously accomplished by requiring 
$3/8n^2<|\Lambda|<3/4n^2$ for positive and negative cosmological. In this case the horizons $r_-$ and $r_{--}$ do not emerge. Now, for the horizons $r_-$ and $r_{--}$ to exist it is observed that the mass parameter needs to satisfy $0<|M|<\frac{\bar{l}}{4}-\frac{n^2}{\bar{l}}$. The conditions $\bar{l}\in\mathbb{R}$ and $\frac{\bar{l}}{4}-\frac{n^2}{\bar{l}}>0$ are then satisfied by 
$0<|\Lambda|<3/8n^2$ for positive and negative cosmological constant. Here, both horizons $r_+$ and $r_{++}$ do not emerge. \\
\\
In the next subsections the cosmological constant is accordingly subjected to these conditions. We explicitly determine the existence or absence of Killing horizons controlling the mass parameter only.

\subsubsection{$\Lambda>0$ and $K=1$}

The curvature singularities depend on the mass parameter and on the values the combination $ab$ can take. We tackle first the $r_0^+$ cases.\\
\\
$r_0^+$: When the curvature singularity $r_0^+$ takes place the radial coordinate assumes positive values in the range $(r_0^+,\infty)$. Hence, according to (\ref{poscuv}) three cases arise.\\
\\
Case I: Here, the mass parameter is positive and it is bounded from below $M>\frac{abn}{\sqrt{1-a^2b^2}}$, with $0<ab<1$. This implies that the pole of the conformal factor located at $r=M$ satisfies $r=M>r_0^+=M-ab\sqrt{n^2+M^2}$. Hence, the two regions $\mathcal{R}^{-}\in(r^{+}_0,M)$ and $\mathcal{R}^{+}\in(M,\infty)$ emerge, dividing the spacetime into two causally disconnected regions with different geometric properties according with the relative position of the Killing horizons. The number of possible Killing horizons depends on the mass parameter. The most generic case corresponds to $\frac{abn}{\sqrt{1-a^2b^2}}<M<\frac{\bar{l}}{4}-\frac{n^2}{\bar{l}}$ which ensures the existence of the four Killing horizons (\ref{horizons}), three of them in the positive range of the radial coordinate. These obey $r_0^+<r_{-}<M<r_{+}<r_{++}$. The interpretation of each of the regions proceeds as follows:
\vspace{0.2cm}
\\
$\mathcal{R}^{+}$: This region is limited by the boundaries $r=M$ and $r=\infty$, whose constant curvatures are given by (\ref{ctecuv}). Two Killing horizons are present, the event horizon $r_{+}$ and the cosmological horizon $r_{++}$. The spacetime region lying in between is recognized as the exterior of a regular stationary black hole, with no pole for the scalar field neither a curvature singularity. This is a consequence of the presence of the new boundary at $r=M$ that forbids the access to $\mathcal{R}^{-}$ where $r_{0}^{+}$ lies. The causal structure of this black hole resembles the one encountered in the Reissner-Nordstr\"om de Sitter black hole but with the timelike singularity replaced by the new asymptotic region at  $r=M$ and with a different distribution of the Killing horizons, see Fig. 1. Instead of the appearance of a cosmological, event and Cauchy horizons, only the first two emerge. For instance, the outer region VI and the inner region I that is found when the cosmological horizon is crossed are identical to the ones found for the Reissner-Nordstr\"om de Sitter black hole. Nevertheless, once the event horizon of solution (\ref{BB1}) is crossed an observer will encounter the new asymptotic region at $r=M$ instead of the region lying between the event and Cauchy horizons that would face in the Reissner-Nordstr\"om de Sitter geometry.
In the context of regular black holes this is an interesting feature, Cauchy horizons are known to be unstable \cite{Maeda:2005yd} or for suffering from mass inflation \cite{Dokuchaev:2013uda}. It is then appealing to investigate how  the event horizon behaves in this solution and to test how robust it is under perturbations. \\
\\
\begin{figure}[h!]
    \includegraphics[width=0.8\textwidth]{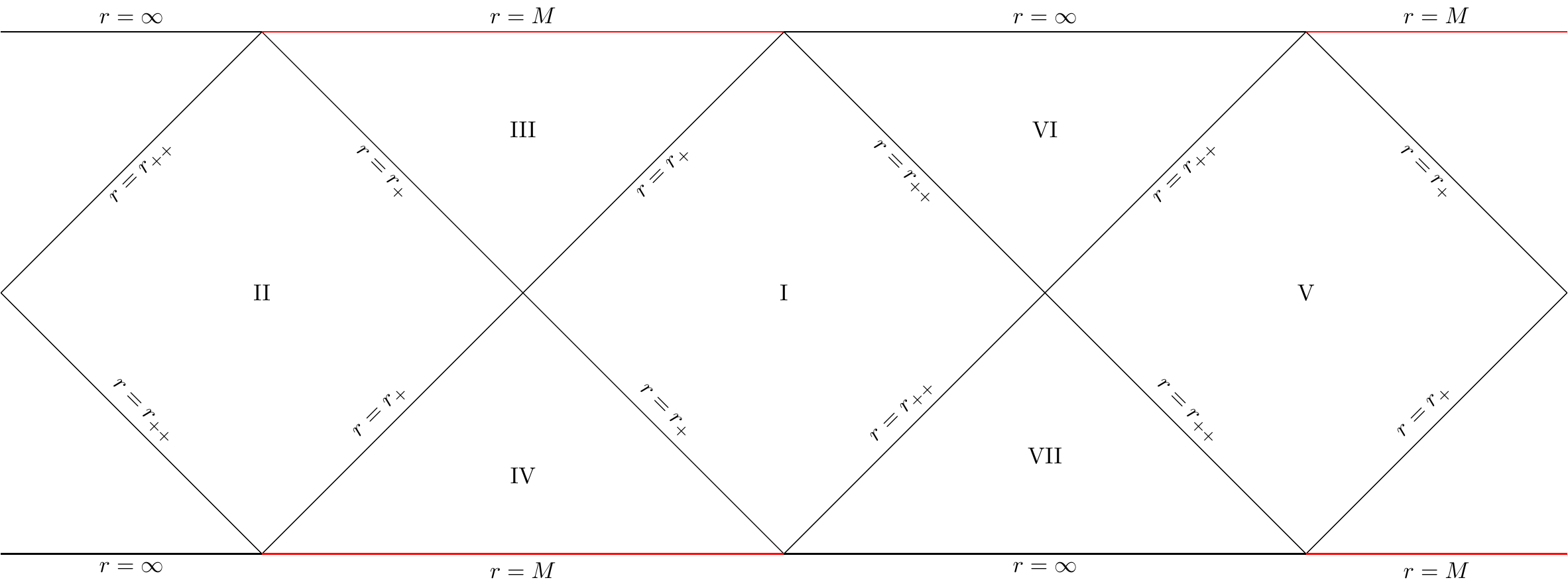}
    \caption{Penrose diagram of the de Sitter regular black hole with event and cosmological horizons contained in solution (\ref{BB1}). The new asymptotic region $r=M$ is specially highlighted.} 
     \label{fig1}
\end{figure}
\\
Now, for $M>\frac{\bar{l}}{4}-\frac{n^2}{\bar{l}}$, $r_{+}$ and $r_{++}$ do not emerge and $\mathcal{R}^{+}$ is devoid of Killing horizons, indeed $r_-$ belongs to $\mathcal{R}^{-}$. The metric is everywhere regular and the scalar field has no pole in contrast with the seed counterpart. Due to the absence of horizons the metric function $F(r)$ is always negative, the $r$ coordinate is timelike and so the spacetime is cosmological. This case is particularly interesting and it offers an explicit example in which the effect of the transformation is notorious. It transforms the seed solution, which behaves as a cosmological model outside of $r_-$ only, into a fully regular bouncing cosmology with no horizons whatsoever and with no divergences for the scalar field. 
This can be seen by investigating the two dimensional determinant for $t=r=constant$ surfaces
\begin{equation}
g_2=\Omega(r)^2(r^2+n^2)\left[(r^2+n^2)\sin^2\theta-16n^2F(r)\sin^4\frac{\theta}{2}\right],
\end{equation}
which provides a measurement of the 2-dimensional volume of the transverse manifold. Due to the fact that $F(r)$ is everywhere negative this volume never vanishes, no matter the values of the radial and polar coordinates. This provides a bouncing cosmology that interpolates between two de Sitter Universes with cosmological constants given by $\frac{\Lambda}{3a^2b^2}$ and $\frac{\Lambda}{3}$, respectively (\ref{ctecuv}). The location of the bounce corresponds to a minimum of $g_2$ which can be graphically observed to occur for an open set of parameter $\Lambda$, $M$ and $n$, such that all the conditions that defines this spacetime are satisfied, see Fig. 2. 
\begin{figure}[h!]
    \includegraphics[width=1.01\textwidth]{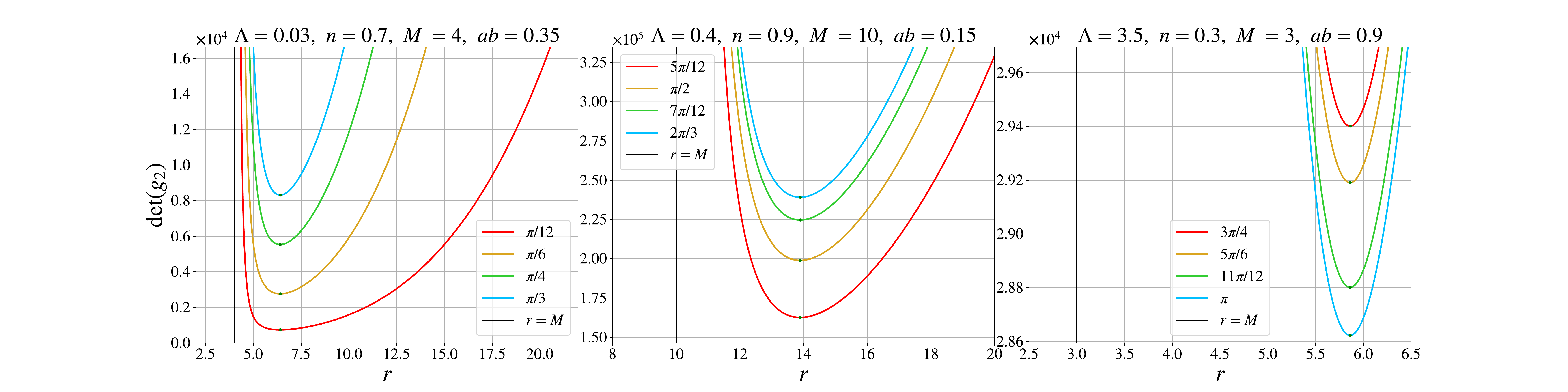}
    \caption{Position of the bounce for different set of parameters $\Lambda$, $M$ and $n$ and for several values of the polar coordinate $\theta$. $r_0^+$ Case I.}
    \label{fig2}
\end{figure}
\\
Let us now notice that the mass parameter might saturate the previous inequalities, namely, $M=\frac{\bar{l}}{4}-\frac{n^2}{\bar{l}}$. For such scenario the cosmological and event horizons merge and only one cosmological horizon $r_+=r_{++}:=r_c$ takes place. This case is particularly interesting, it might define either a de Sitter wormhole that connects the two asymptotic regions $r=M$ and $r=\infty$ when the minimum of $g_2$, the throat, is located behind the cosmological horizon $r_c$ or a bouncing cosmology connecting two de Sitter Universes when the minimum of $g_2$, in this case the bounce, is located outside the cosmological horizon. The bouncing cosmology is clearer, $F(r)$ is always negative outside the cosmological horizon an so the bounce is guaranteed. However, for the wormhole case the situation is less trivial due to the fact that inside the cosmological horizon $F(r)$ is actually always positive. Nonetheless, 
in the space parameter that defines this solution it is possible to find a set in which the nontrivial minimum volume is preserved in the wormhole case as well. Both cases are graphically shown to occur, see Fig. 3.
\begin{figure}[h!]
    \includegraphics[width=1.01\textwidth]{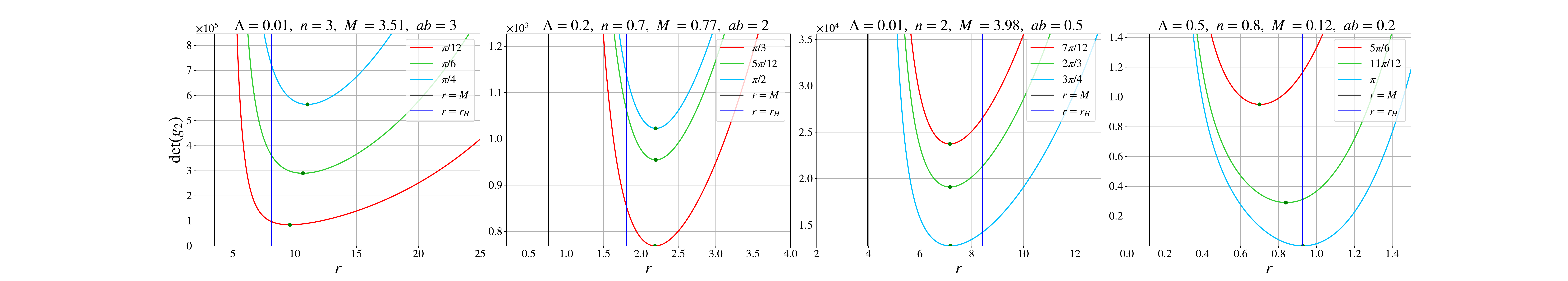}
    \caption{The first two diagrams on the left hand side represent locations of the bounces for the bouncing cosmology. On the other hand, the next two diagrams on the right hand side represent positions for the wormhole throat. In both cases different values of the parameters $\Lambda$, $M$ and $n$ and several values of the polar coordinate $\theta$ are shown. The location of the cosmological horizon is explicitly indicated.}
    \label{fig3}
\end{figure}
The conformal diagram of these solutions is given in Fig. 4.
\begin{figure}[h!]
    \includegraphics[width=0.8\textwidth]{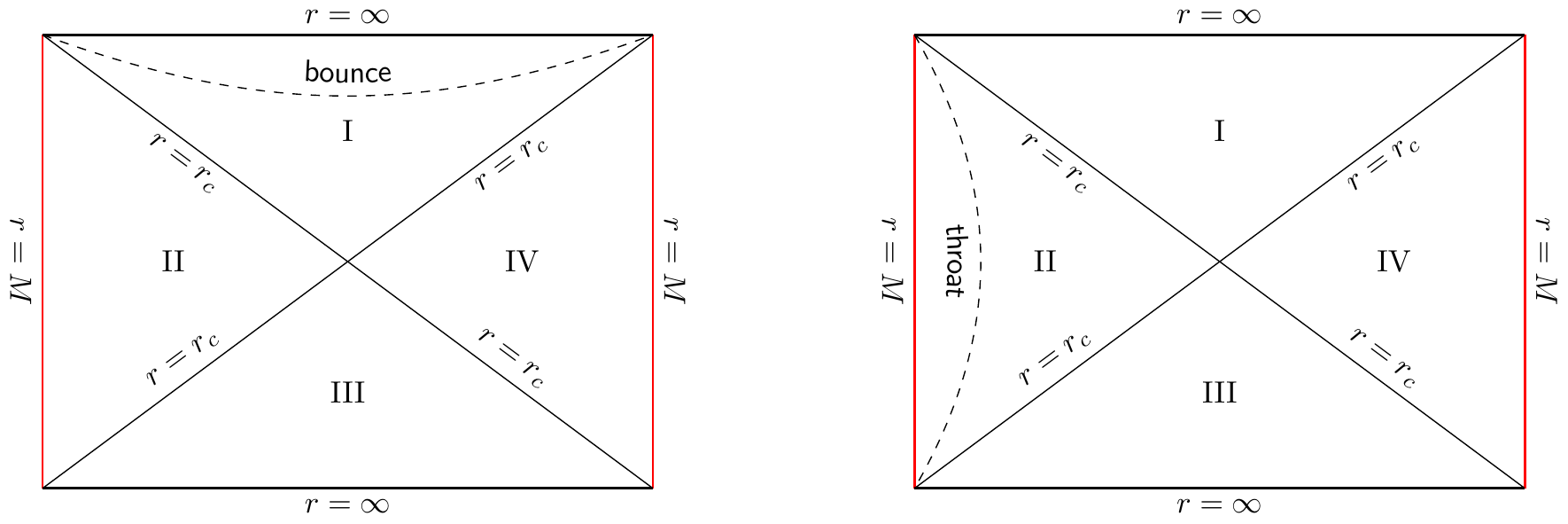}
    \caption{Conformal diagrams for the bouncing cosmology and the de Sitter wormhole configurations. The asymptotic region $r=M$, and the bounce and throat locations are specially highlighted.}
    \label{fig4}
\end{figure}
The diagram on the left side shows the conformal structure of the bouncing cosmology. The bounce, represented by the dashed line, connects the time dependent region I with the stationary region II. Once the bounce is surpassed an observer will encounter the cosmological horizon $r_c$, which after being crossed gives access to the stationary region II, latter bounded by the asymptotic region $r=M$. On the other hand, the diagram on the right side shows the de Sitter wormhole. The throat is located in region II and connects the two asymptotic boundaries at $r=\infty$ and $r=M$.\\
\\
Finally, region $\mathcal{R}^+$ represents an exact de Sitter black bounce which smoothly transits from a regular stationary black hole to a bouncing cosmology that interpolates between two de Sitter Universes. An intermediate phase represented by two different configurations is observed. A bouncing cosmology or a de Sitter wormhole, both connecting the asymptotic regions $r=M$ and $r=\infty$.
\begin{equation} dS\hspace{0.1cm}black\hspace{0.1cm}bounce \rightarrow
\left\{ \begin{aligned} 
\frac{abn}{\sqrt{1-a^2b^2}}<&M<\frac{\bar{l}}{4}-\frac{n^2}{\bar{l}}\hspace{0.2cm} \rightarrow \hspace{0.3cm} regular\hspace{0.1cm}stationary\hspace{0.1cm}black\hspace{0.1cm}hole\\
&M=\frac{\bar{l}}{4}-\frac{n^2}{\bar{l}}\hspace{0.2cm} \rightarrow \hspace{0.3cm}bouncing\hspace{0.1cm}cosmology\hspace{0.1cm}or\hspace{0.1cm}wormhole\\
\frac{\bar{l}}{4}-\frac{n^2}{\bar{l}}<&M<\infty \hspace{1cm} \rightarrow  \hspace{0.3cm}bouncing\hspace{0.1cm}cosmology.
\end{aligned} \right. \label{BB1}
\end{equation}
\\
$\mathcal{R}^{-}$: This region lies between $r_0^{+}$ and the asymptotic region $r=M$. Only the cosmological horizon $r_{-}$ might be present when $M<\frac{\bar{l}}{4}-\frac{n^2}{\bar{l}}$. Two cases arise, $r_0^{+}<r_-$ and $r_0^{+}>r_-$, both representing naked singularities.\\
\\
Case II: This case is defined by $M>0$ and $ab<0$. The curvature singularity satisfies $r_0^+=M+|ab|\sqrt{n^2+M^2}>r=M$ and in consequence the conformal factor pole does not belong to the domain of the radial coordinate $(r_0^+,\infty)$. Hence, only one region takes place which might be dressed by up to three Killing horizons depending on whether or not the mass parameter satisfies $0<M<\frac{\bar{l}}{4}-\frac{n^2}{\bar{l}}$. The location of the curvature singularity with respect to the Killing horizons opens three possible cases: (i) $r_0^+<r_+<r_{++}$, (ii) $r_0^+<r_{++}$, and (iii) $r_{++}<r_0^+$. It is always observed that $r_-<r_0^+$. (i) represents a black hole with two horizons, the event horizon $r_+$ and the cosmological horizon $r_{++}$. (ii) exhibits only one horizon, the cosmological horizon $r_{++}$, while (iii) is free of horizons. The last two cases correspond to spacetimes with naked singularities. Finally, for $M>\frac{\bar{l}}{4}-\frac{n^2}{\bar{l}}$ no Killing horizons appear and all cases represent naked singularities. \\
\\
Case III: This case possesses a negative mass satisfying $|M|<\frac{|ab|n}{\sqrt{1-|ab|^2}}$, provided that $0<|ab|<1$ holds. Once again, the conformal factor diverges outside of the domain of the radial coordinate $r=-|M|<r_0^+=-|M|+|ab|\sqrt{n^2+M^2}$. 
We rewrite the Killing horizons by performing the change $M\rightarrow-|M|$ on (\ref{horizons})
\begin{align}
r_{--}&=\frac{\bar{l}}2\left( -1-\sqrt{1-4\frac{n^2}{\bar{l}^2}-4\frac{|M|}{\bar{l}}} \right)\\
r_{-}&=\frac{\bar{l}}2\left( -1+\sqrt{1-4\frac{n^2}{\bar{l}^2}-4\frac{|M|}{\bar{l}}} \right) \\
r_+&=\frac{\bar{l}}2\left( 1-\sqrt{1-4\frac{n^2}{\bar{l}^2}+4\frac{|M|}{\bar{l}}} \right)\\
r_{++}&= \frac{\bar{l}}2\left( 1+\sqrt{1-4\frac{n^2}{\bar{l}^2}+4\frac{|M|}{\bar{l}}} \right).  \label{horizons2}
\end{align}
It is observed that the Killing horizons $r_{--}$ and $r_-$ always lie outside of the domain $(r_0^+,\infty)$. Thus, the spacetime contains a curvature singularity at $r_0^+$ which might be covered by up to two horizons $r_+$ and $r_{++}$ as long as $|M|>\frac{n^2}{\bar{l}}-\frac{\bar{l}}{4}$. 
Two cases arise: (i) $r_0^+<r_+<r_{++}$ and (ii) $r_0^+<r_{++}$. Again, the first case represents a stationary black hole with two horizons, the event horizon $r_+$ and the cosmological horizon $r_{++}$. On the other hand, the second case exhibits only the cosmological horizon and the singularity remains naked. Naked singularities are obtained for $0<|M|<\frac{n^2}{\bar{l}}-\frac{\bar{l}}{4}$. \\
\\
$r_0^-$: When the curvature singularity $r_0^-$ takes place the radial coordinate assumes negative and positive values in the range $(r_0^-,\infty)$. Hence, according to (\ref{negcuv}) three cases arises.\\
\\
Case I: This case is defined by $0<M<\frac{abn}{\sqrt{1-a^2b^2}}$ with $0<ab<1$. The pole of the conformal factor satisfies $r=M>r_0^-=M-ab\sqrt{n^2+M^2}$, and in consequence belongs to the allowed domain of the radial coordinate. The spacetime is thus divided into the two regions $\mathcal{R}^{-}\in(r^{-}_0,M)$ and $\mathcal{R}^{+}\in(M,\infty)$, which are again causally disconnected. All Killing horizons exist for $M<\frac{\bar{l}}{4}-\frac{n^2}{\bar{l}}$. In contrast to the first case in (\ref{poscuv}), here the radial coordinate also takes negative values and in principle up to four Killing horizons might take part in the causal structure of this solution. However, it is observed that only $r_{-}$, $r_{+}$, and $r_{++}$ belong to the domain of the radial coordinate.
For each of the regions we have:
\vspace{0.2cm}
\\
$\mathcal{R}^{+}$: This region is composed by the boundaries $r=M$ and $r=\infty$. The event $r_+$ and the cosmological $r_{++}$ horizons are englobed, and a regular stationary black hole with a similar horizon structure to the one of the Reissner-Nordstr\"om de Sitter black hole is again obtained. The causal structure is fairly represented by the conformal diagram of Fig. 1. 
Notice that in this case to make $r_+$ and $r_{++}$ disappear the mass parameter must satisfies $\frac{\bar{l}}{4}-\frac{n^2}{\bar{l}}<M<\frac{abn}{\sqrt{1-a^2b^2}}$. A new bouncing cosmology similar to the one previously reported is obtained in this case as well, now in a different subset of the parameter space. See Fig. 5. Again, the two configurations explained in the previous black bounce are obtained when saturating the value of the mass parameter. We do not provide a new plot for this case.
The following exact de Sitter black bounce is then identified
\begin{equation} dS\hspace{0.1cm}black\hspace{0.1cm}bounce \rightarrow
\left\{ \begin{aligned} 
0<&M<\frac{\bar{l}}{4}-\frac{n^2}{\bar{l}}\leq\frac{abn}{\sqrt{1-a^2b^2}}\hspace{0.33cm} \rightarrow \hspace{0.3cm} regular\hspace{0.1cm}stationary\hspace{0.1cm}black\hspace{0.1cm}hole\\
&M=\frac{\bar{l}}{4}-\frac{n^2}{\bar{l}}\hspace{2.4cm} \rightarrow \hspace{0.3cm}bouncing\hspace{0.1cm}cosmology\hspace{0.1cm}or\hspace{0.1cm}wormhole\\
\frac{\bar{l}}{4}-\frac{n^2}{\bar{l}}<&M<\frac{abn}{\sqrt{1-a^2b^2}} \hspace{1.9cm} \rightarrow  \hspace{0.3cm}bouncing\hspace{0.1cm}cosmology.
\end{aligned} \right. 
\end{equation}
\vspace{0.2cm}
\\
\begin{figure}[h!]
    \includegraphics[width=1.01\textwidth]{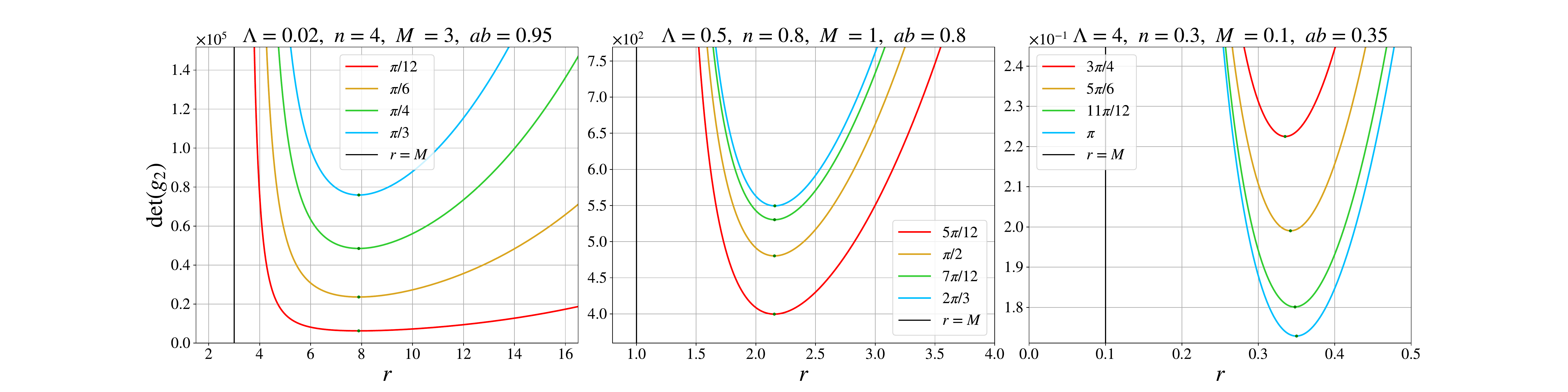}
    \caption{Position of the bounce for different sets of parameters $\Lambda$, $M$ and $n$ and for several values of the polar coordinate $\theta$, now for Case I contained in $r_0^-$.}
    \label{fig5}
\end{figure}
\\
$\mathcal{R}^{-}$: This region lies between $r_0^{-}$ and the asymptotic region $r=M$. Two cases arise: $r_0^-<r_-<0<M$ and $r_0^-<0<r_-<M$. In both scenarios the curvature singularity is covered by the  cosmological horizon $r_-$, then observers outside $r_-$ are not protected from the singularity.\\
\\
Case II: The mass parameter satisfies $|M|>\frac{|ab|n}{\sqrt{1-|ab|^2}}$, where $0<|ab|<1$. The pole of the conformal factor is situated at $r=-|M|$, and it satisfies $r=-|M|<r_0^-=-|M|+|ab|\sqrt{n^2+|M|^2}$. It does not belong to the domain $(r_0^-,\infty)$ and no split of the spacetime occurs. Only the horizons $r_+$ and $r_{++}$ take place as soon as the mass parameter obeys $\frac{|ab|n}{\sqrt{1-|ab|^2}}\leq\frac{n^2}{\bar{l}}-\frac{\bar{l}}{4}<|M|$. 
Hence, the following cases  arise: A stationary black hole with a curvature singularity covered by the event $r_+$ and cosmological $r_{++}$ horizons, where $r_0^-<r_+<0<r_{++}$ or $r_0^-<0<r_+<r_{++}$ and a curvature singularity covered by the cosmological horizon $r_{++}$ only, the latter representing a naked singularity. For $\frac{|ab|n}{\sqrt{1-|ab|^2}}<|M|<\frac{n^2}{\bar{l}}-\frac{\bar{l}}{4}$ both horizons disappear and only naked singularities take place. \\
\\
Case III: The mass parameter is negative $M<0$ and $ab>0$. In this case the pole of the conformal factor satisfies $r=-|M|>r_0^-=-|M|-ab\sqrt{n^2+M^2}$, belonging to the domain $(r_0^-,\infty)$ and thus dividing the spacetime into the regions $\mathcal{R}^{-}\in(r^{-}_0,-|M|)$ and $\mathcal{R}^{+}\in(-|M|,\infty)$. 
Each region proceeds as follows.
\vspace{0.2cm}
\\
$\mathcal{R}^{+}$: This region is limited by $r=-|M|$ and $r=\infty$. For values of the mass such that $|M|>\frac{n^2}{\bar{l}}-\frac{\bar{l}}{4}$ the two Killing horizons $r_+$ and $r_{++}$ emerge, which satisfy  either $-|M|<0<r_+<r_{++}$ or $-|M|<r_+<0<r_{++}$. The spacetime contained in between the event and cosmological horizons is recognized once again as the exterior of a stationary regular black hole with a similar structure to the one of the Reissner-Nordstr\"om de Sitter geometry, see Fig. 2. Moving the mass parameter such that $0<|M|<\frac{n^2}{\bar{l}}-\frac{\bar{l}}{4}$ is achieved, then both horizons disappear and once again a bouncing cosmology interpolating between two de Sitter Universes is obtained. See Fig. 6. Saturating the mass inequality the bouncing cosmology or the de Sitter wormhole previously explained emerge. Ultimately, the spacetime describes the de Sitter black bounce 
\begin{equation} dS\hspace{0.1cm}black\hspace{0.1cm}bounce \rightarrow
\left\{ \begin{aligned} 
\frac{n^2}{\bar{l}}-\frac{\bar{l}}{4}<&|M|<\infty\hspace{1cm} \rightarrow \hspace{0.3cm} regular\hspace{0.1cm}stationary\hspace{0.1cm}black\hspace{0.1cm}hole\\
&|M|=\frac{n^2}{\bar{l}}-\frac{\bar{l}}{4}\hspace{0.2cm} \rightarrow \hspace{0.3cm} bouncing\hspace{0.1cm}cosmology\hspace{0.1cm}or\hspace{0.1cm}wormhole\\
0<&|M|<\frac{n^2}{\bar{l}}-\frac{\bar{l}}{4} \hspace{0.25cm} \rightarrow  \hspace{0.3cm}bouncing\hspace{0.1cm}cosmology.
\end{aligned} \right. 
\end{equation}
\\
\begin{figure}[h!]
    \includegraphics[width=1.01\textwidth]{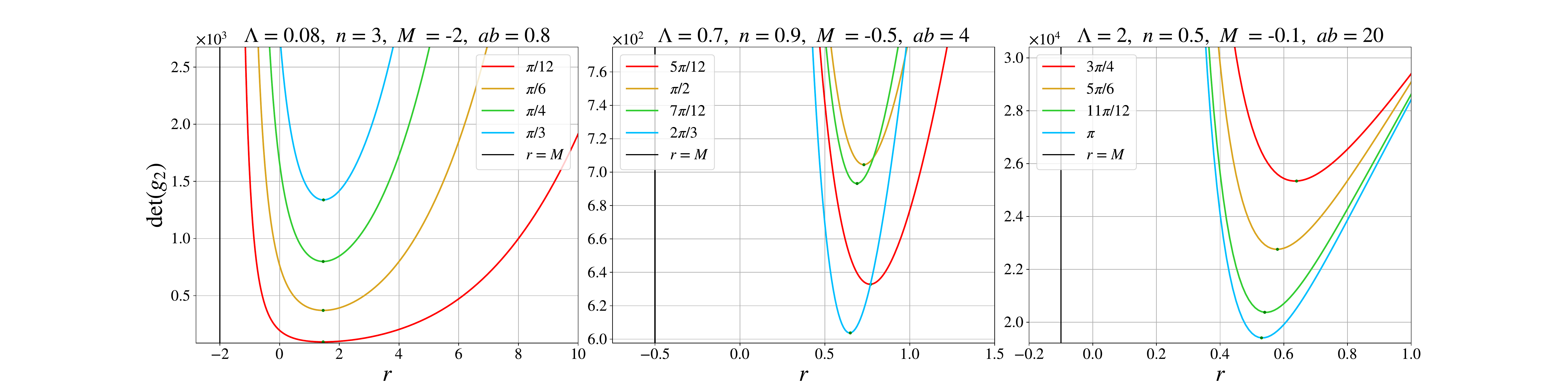}
    \caption{Position of the bounce for different set of parameters $\Lambda$, $M$ and $n$ and for several values of the polar coordinate $\theta$, now for Case III contained in $r_0^-$.}
    \label{fig6}
\end{figure}\\
\\
$\mathcal{R}^{-}$: This region is limited by $r=r_0^-$ and $r=-|M|$. For a mass parameter $|M|<\frac{\bar{l}}{4}-\frac{n^2}{\bar{l}}$ the two Killing horizons $r_{--}$ and $r_{-}$ take place defining two possible cases: (i) $r_0^-<r_{--}<r_-<-|M|$ and (ii) $r_0^-<r_-<-|M|$.
In the first case the region between the event horizon $r_{--}$ and the cosmological horizon $r_{-}$ defines the exterior of a stationary black hole, while on the other hand the second case in which only $r_-$ is present defines a naked singularity.

\subsubsection{$\Lambda<0$ and $K=-1$}

$r_0^+$: The radial coordinate assumes positive values in the range $(r_0^+,\infty)$. We consider each of the cases contained in (\ref{poscuv}). We do not repeat here all constraints on $M$ and $ab$ regarding the position of this singularity.\\
\\
Case I: The pole of the conformal factor satisfies $r=M>r_0^+=M-ab\sqrt{n^2+M^2}$. The spacetime splits into the regions $\mathcal{R}^{-}\in(r^{+}_0,M)$ and $\mathcal{R}^{+}\in(M,\infty)$. For mass parameter values such that $0<M<\frac{\bar{l}}{4}-\frac{n^2}{\bar{l}}$ three horizons in the domain of the radial coordinate take place, $r_0^+<r_{-}<M<r_{+}<r_{++}$. These horizons are now free of cosmological behavior, the cosmological constant is negative.
\vspace{0.2cm}
\\
$\mathcal{R}^{+}$: This region is limited by $r=M$ and $r=\infty$ and it encloses the inner and event horizons $r_+$ and $r_{++}$. Again, the asymptotic region $r=M$ replaces the singularity and provides a regular metric with no poles for the scalar field. The spacetime is interpreted as a regular stationary black hole with a similar causal structure to the one of the Reissner-Nordstr\"om anti--de Sitter black hole. Its conformal diagram is depicted in Fig. 7.
\\
\begin{figure}[h!]
    \includegraphics[width=0.3\textwidth]{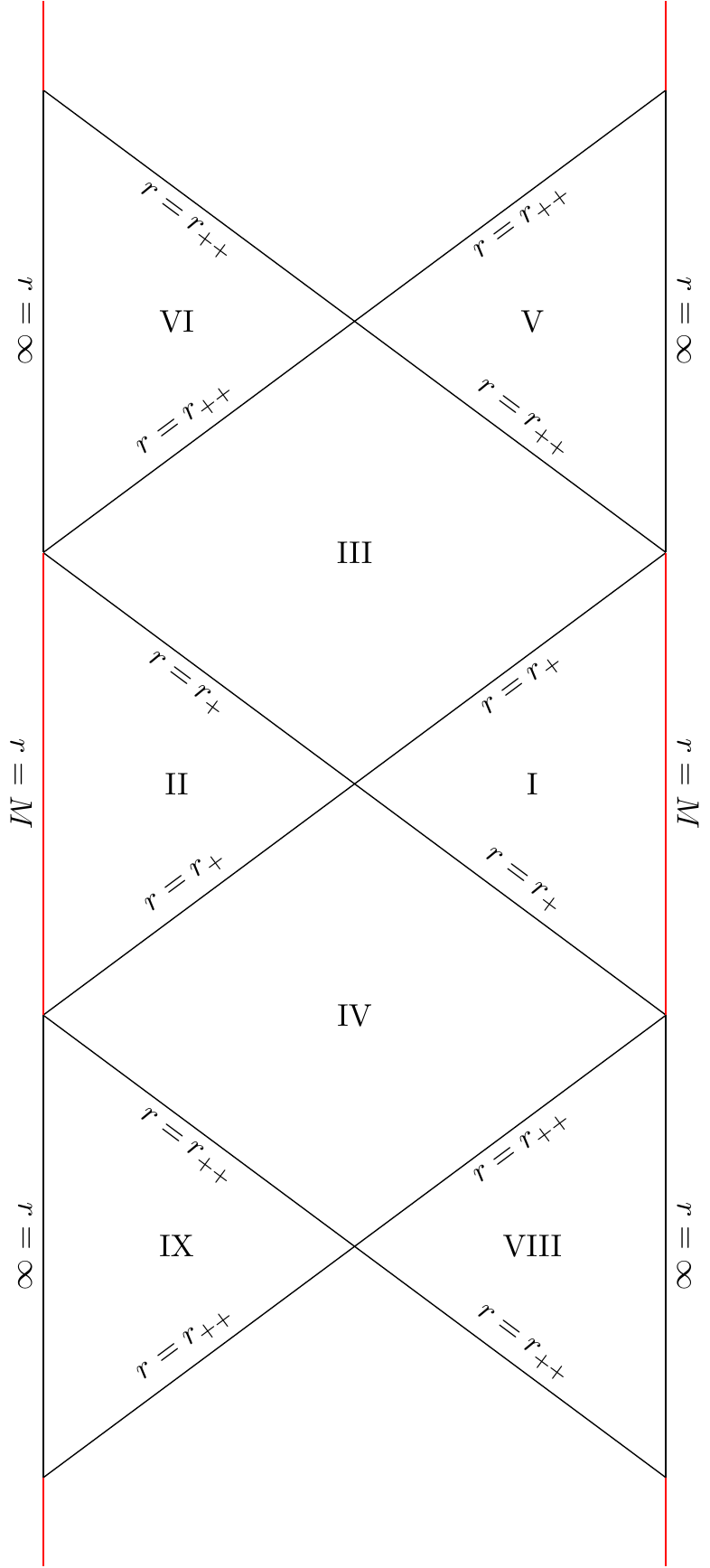}
    \caption{Penrose diagram of the anti--de Sitter regular black hole with inner and event horizons contained in solution (\ref{BB4}). The new asymptotic region $r=M$ is specially highlighted.}
    \label{fig7}
\end{figure}
\\
From Fig. 7 it is again observed how the asymptotic region at $r=M$ replaces the singularity. This diagram does not radically differ from the one describing the de Sitter case. The main differences are two: First, the outer region is now stationary, $r_{++}$ plays the role of an event horizon. Second, the inner horizon $r_+$ appears beyond which the asymptotic region is encountered. \\
\\
Particularly appealing is the case in which the mass obeys the bound $M=\frac{\bar{l}}{4}-\frac{n^2}{\bar{l}}$, for which both horizons merge in a single event horizon $r_+=r_{++}:=r_H$. 
It is observed that $r_H$ takes place at the right-hand side of the asymptotic region $r=M$ and so it belongs to $\mathcal{R}^+$. This occurs for several open sets of the corresponding parameters $\Lambda$, $M$, and $n$. Therefore, two possible configurations might emerge. The first possible configuration corresponds to a regular anti--de Sitter black hole with a single event horizon which lies inside of an anti--de Sitter wormhole. On the other hand, the second configuration represents a bouncing cosmology lying at the interior of a regular anti--de Sitter black hole with a single event horizon. The location of the bounce for the latter configuration is trivial to obtain, namely, $F(r)$ is everywhere negative inside the event horizon $r_H$. 
Conversely, due to the positivity of $F(r)$ outside $r_H$ the appearance of the wormhole throat is less trivial, see the discussion below Eq. (\ref{dethip}).
Nonetheless, in Fig. 8 we show the existence of both, the bounce and the throat, for different sets of the parameters $\Lambda$, $M$, and $n$. 
\begin{figure}[h!]
    \includegraphics[width=1.01\textwidth]{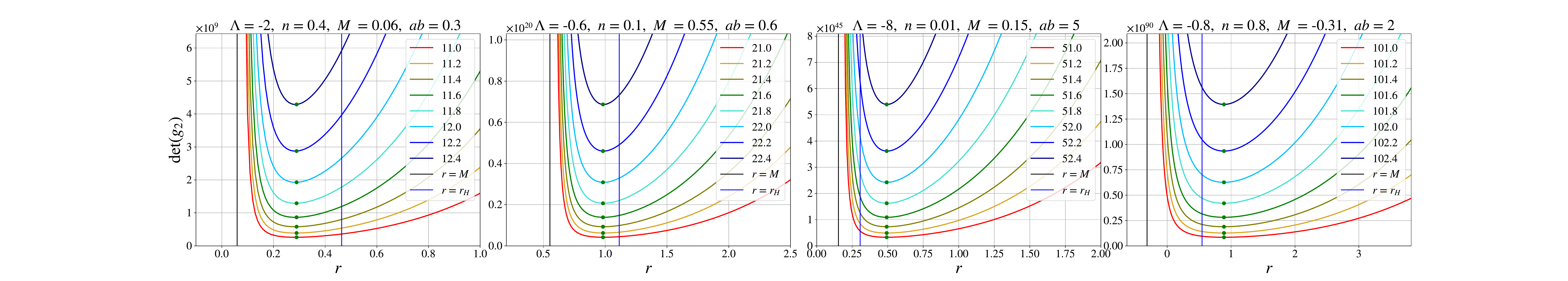}
    \caption{The first two diagrams in the left hand side represent locations for the bounce of the bouncing cosmology inside the anti--de Sitter black hole. On the other hand, the next two diagrams in the right hand side represent the position of the wormhole throat inside of which lies the anti--de Sitter black hole. In both cases different values of the parameters $\Lambda$, $M$, and $n$ and several values of the $\theta$ coordinate are shown. The location of the event horizon is explicitly indicated.}
    \label{fig8}
\end{figure}
\\
The conformal diagram of these solutions is depicted in Fig. 9. For both cases the general picture is very similar and it resembles the causal structure encountered in the Schwarzschild anti--de Sitter black hole however with the central singularity replaced by the asymptotic region $r=M$. This converts the solution into a fully regular black hole with one single event horizon. The latter is a very exciting feature for a regular black hole, indeed most of regular black hole solutions include Cauchy horizons as well which are known for being subjected to different types of instabilities. This solution represents an excellent candidate for a fully regular black hole with no perturbative instabilities. The diagram on the left-hand side of Fig. 9 shows the case in which the regular black hole is immersed in an anti--de Sitter wormhole. Thus, an observer coming from asymptotic infinity in region I first observe a wormhole which once is crossed leads to an anti--de Sitter regular black hole. Region II represents the interior of this black hole where no singularity takes place. Such an observer stays trapped in region I. The diagram on the right-hand side of Fig. 9 represents the same regular black hole but with a cosmological bounce in its interior. Thus, after the observer has crossed the event horizon the interior of the black hole is given by a bouncing cosmology that leads to the asymptotic region $r=M$, a region from which the observer cannot escape.\\
\begin{figure}[h!]
    \includegraphics[width=0.8\textwidth]{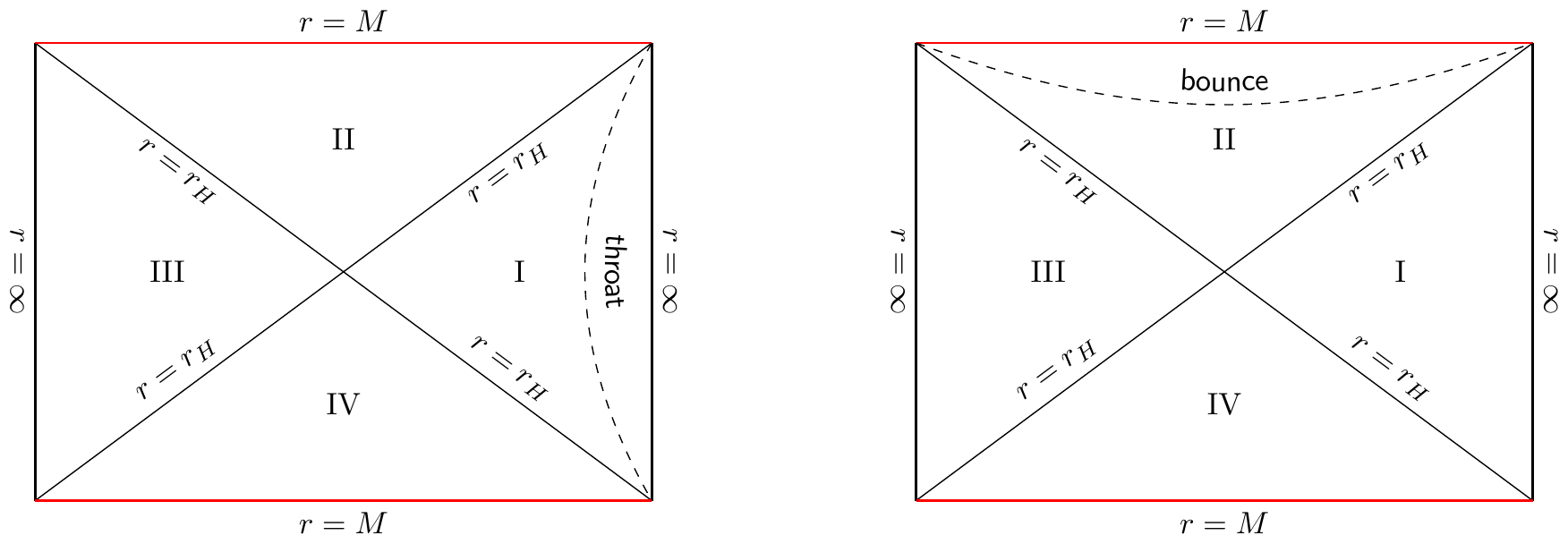}
    \caption{Conformal diagrams for the bouncing cosmology inside of the anti--de Sitter regular black hole and for the anti--de Sitter regular black hole inside of the anti--de Sitter wormhole. The asymptotic region $r=M$, and the bounce and throat locations are especially highlighted.}
    \label{fig9}
\end{figure}
\\ 
Now, as for the de Sitter solutions previously analyzed it is desirable to understand what occurs for $M>\frac{\bar{l}}{4}-\frac{n^2}{\bar{l}}$. In such a case both horizons $r_+$ and $r_{++}$ disappear and the spacetime represents a regular gravitational soliton. The metric function is everywhere positive and the spacetime remains stationary. 
In this case the effect of the transformations is also decisive. The two dimensional determinant $g_2$ is now given by
\begin{equation}
g_2=\Omega(r)^2(r^2+n^2)\left[(r^2+n^2)\sinh^2\theta-16n^2F(r)\sinh^4\frac{\theta}{2}\right].    \label{dethip}
\end{equation}
Contrary to the cases with spherical foliation, the metric function $F(r)$ is everywhere positive and in principle $g_2$ may vanish and even more take negative values. A negative $g_2$ is a consequence of the appearance of closed timelike curves as can be seen from 
\begin{equation}
\bar{g}_{\varphi\varphi}=\Omega(r)\left[(r^2+n^2)\sinh^2\theta-16n^2F(r)\sinh^4\frac{\theta}{2}\right].  \label{CTCsol}
\end{equation}
In spite of this, $\bar{g}_{\varphi\varphi}$ will always remain positive if the cosmological constant is restricted to $|\Lambda|<3/4n^2$ (see the seed case), implying at the same time a nontrivial minimum for the 2-dimensional volume of the transverse manifold for all values of the coordinates $r$ and $\theta$.
Our spacetime then represents a hyperbolic wormhole connecting two asymptotically locally anti--de Sitter boundaries at $r=M$ and $r=\infty$ with asymptotic constant curvatures given by (\ref{ctecuv}). The throat of the wormhole can be graphically observed for a wide set of parameters, see Fig. 10.
\begin{figure}[h!]
    \includegraphics[width=1.01\textwidth]{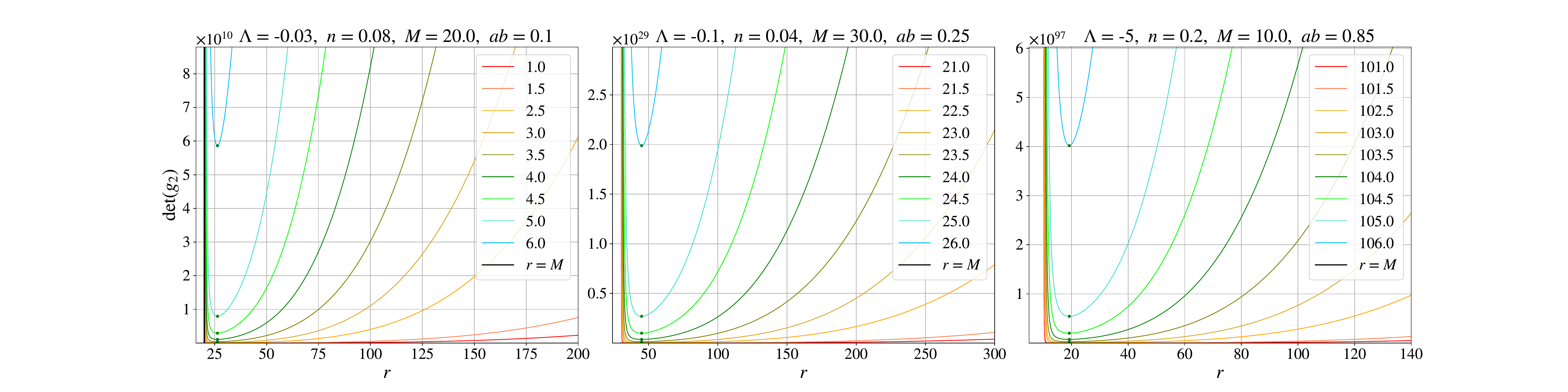}
    \caption{Position of the wormhole throat for different set of parameters $\Lambda$, $M$ and $n$ and for several values of the coordinate $\theta$.}
    \label{fig10}
\end{figure}
\\
Finally, region $\mathcal{R}^{+}$ provides an exact anti--de Sitter black bounce that interpolates between a regular black hole with inner $r_+$ and event $r_{++}$ horizons and a wormhole. The intermediate phase is given by a regular black hole with a single event horizon $r_H$ which might appear in two different configurations, inside an anti--de Sitter wormhole or with an interior given by a cosmological bounce. 
Again, all these transitions are smoothly connected by the mass parameter only.
\begin{equation} AdS\hspace{0.1cm}black\hspace{0.1cm}bounce \rightarrow
\left\{ \begin{aligned} 
0<&M<\frac{\bar{l}}{4}-\frac{n^2}{\bar{l}}\hspace{0.7cm} \rightarrow \hspace{0.3cm} regular\hspace{0.1cm}black\hspace{0.1cm}hole\hspace{0.1cm}with\hspace{0.1cm}inner\hspace{0.1cm}and\hspace{0.1cm}event\hspace{0.1cm}horizons\\
&M=\frac{\bar{l}}{4}-\frac{n^2}{\bar{l}} \hspace{0.66cm} \rightarrow  \hspace{0.3cm}regular\hspace{0.1cm}black\hspace{0.1cm}hole\hspace{0.1cm}with\hspace{0.1cm}one\hspace{0.1cm}event\hspace{0.1cm}horizon\\
&M>\frac{\bar{l}}{4}-\frac{n^2}{\bar{l}} \hspace{0.66cm} \rightarrow  \hspace{0.3cm}wormhole.
\end{aligned} \right. \label{BB4}
\end{equation}
\vspace{0.2cm}
\\
$\mathcal{R}^{-}$: This region is bounded by $r_0^{+}$ and the asymptotic region $r=M$. For mass parameter satisfying $0<M<\frac{\bar{l}}{4}-\frac{n^2}{\bar{l}}$ the singularity is covered by the event horizon $r_-$. The spacetime represents an anti--de Sitter black hole with a single event horizon which is locally isomorphic to $\mathcal{H}^2$. 
\\
\\
Case II: Here, the curvature singularity satisfies $r_0^+=M+|ab|\sqrt{n^2+M^2}>r=M$ and then no division of the spacetime is performed. The region $(r_0^{+},\infty)$ might be then dressed by up to three horizons as soon as the mass parameter follows $0<M<\frac{\bar{l}}{4}-\frac{n^2}{\bar{l}}$. The distribution of the inner and event horizons with respect to the singularity opens three cases:
(i) $r_0^+<r_+<r_{++}$, (ii) $r_0^+<r_{++}$, and (iii) $r_{++}<r_0^+$. $r_-<r_0^+$ always holds. (i) represents a black hole with two horizons, the inner horizon $r_+$ and the event horizon $r_{++}$. (ii) provides a black hole with a single horizon, the event horizon $r_{++}$, while (iii) is free of horizons. The last case corresponds to a naked singularity. Finally, for $M>\frac{\bar{l}}{4}-\frac{n^2}{\bar{l}}$ the horizons $r_+$ and $r_{++}$ are not present and $r_0^+$ remains naked.  \\
\\
Case III: In this case we have $r=-|M|<r_0^+=-|M|+|ab|\sqrt{n^2+M^2}$, thus the conformal factor diverges outside of the domain of the radial coordinate. No new asymptotic region takes place. All Killing horizons are given by (\ref{horizons2}). The Killing horizons $r_-$ and $r_{--}$ always lie outside the domain $(r_0^+,\infty)$. For $|M|>\frac{n^2}{\bar{l}}-\frac{\bar{l}}{4}$ two cases arise: (i) $r_0^+<r_+<r_{++}$ and (ii) $r_0^+<r_{++}$. The first case corresponds to a black hole with two horizons, the internal and event horizons $r_+$ and $r_{++}$ while the second case represents a black hole with one single event horizon $r_{++}$. As soon as $0<|M|<\frac{n^2}{\bar{l}}-\frac{\bar{l}}{4}$ both horizons disappear and the singularity remains naked.\\
\\
$r_0^-$: The radial coordinate assumes negative values in the range $(r_0^-,\infty)$. We consider each of the cases contained in (\ref{negcuv}). Again, we do not repeat all constraints on $M$ and $ab$.\\
\\
Case I: Here we have $r=M>r_0^-=M-ab\sqrt{n^2+M^2}$, as a consequence the pole of the conformal factor belongs to the domain of the radial coordinate. Once again our spacetime splits into the causally disconnected regions $\mathcal{R}^{-}\in(r^{-}_0,M)$ and $\mathcal{R}^{+}\in(M,\infty)$. Three Killing horizons might take part of the causal structure, as long as $M<\frac{\bar{l}}{4}-\frac{n^2}{\bar{l}}$. It is always observed that $r_{--}<r_0^-$. Thus:
\vspace{0.2cm}
\\
$\mathcal{R}^{+}$:
The boundaries $r=M$ and $r=\infty$ encompass two Killing horizons, the inner and event horizons $r_+$ and $r_{++}$, giving the spacetime the interpretation of a stationary regular black hole with a similar causal structure to the Reissner-Nordstr\"om anti--de Sitter black hole. Its causal structure can be seen from Fig. 7.
For $M=\frac{\bar{l}}{4}-\frac{n^2}{\bar{l}}$ the aforementioned horizons merge and a black hole with one single horizon $r_H$ appears. Again, this horizon satisfies $r_H>r=M$. The same configurations previously discussed in this context are again present, see Fig. 9.
Finally, for $\frac{\bar{l}}{4}-\frac{n^2}{\bar{l}}<M<\frac{abn}{\sqrt{1-a^2b^2}}$ both horizons disappear and the spacetime represents a wormhole connecting the two locally asymptotically anti--de Sitter boundaries with constant curvatures $\frac{\Lambda}{3a^2b^2}$ and $\frac{\Lambda}{3}$, respectively. The wormhole throat is graphically observable in Fig. 11. This region represents an anti--de Sitter black bounce smoothly connecting, through the mass parameter, a regular black hole with two horizons, a single event horizon black hole and a hyperbolic wormhole.
\begin{equation} AdS\hspace{0.1cm}black\hspace{0.1cm}bounce \rightarrow
\left\{ \begin{aligned} 
0<&M<\frac{\bar{l}}{4}-\frac{n^2}{\bar{l}}\leq\frac{abn}{\sqrt{1-a^2b^2}}\hspace{0.33cm} \rightarrow \hspace{0.3cm} regular\hspace{0.1cm}black\hspace{0.1cm}hole\hspace{0.1cm}with\hspace{0.1cm}inner\hspace{0.1cm}and\hspace{0.1cm}event\hspace{0.1cm}horizons\\
&M=\frac{\bar{l}}{4}-\frac{n^2}{\bar{l}} \hspace{2.4cm} \rightarrow  \hspace{0.3cm}regular\hspace{0.1cm}black\hspace{0.1cm}hole\hspace{0.1cm}with\hspace{0.1cm}one\hspace{0.1cm}event\hspace{0.1cm}horizon\\
\frac{\bar{l}}{4}-\frac{n^2}{\bar{l}}<&M<\frac{abn}{\sqrt{1-a^2b^2}} \hspace{1.9cm} \rightarrow  \hspace{0.3cm} wormhole.
\end{aligned} \right. 
\end{equation}
\begin{figure}[h!]
    \includegraphics[width=1.01\textwidth]{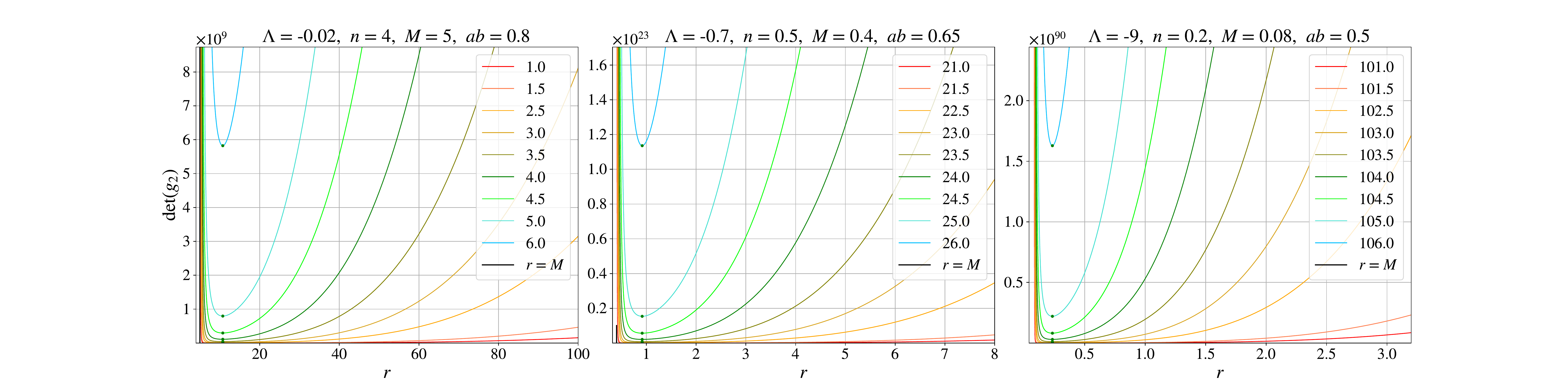}
    \caption{Position of the wormhole throat for different set of parameters $\Lambda$, $M$ and $n$ and for several values of the coordinate $\theta$.}
    \label{fig11}
\end{figure}
\\
\vspace{0.2cm}
$\mathcal{R}^{-}$: For $0<M<\frac{\bar{l}}{4}-\frac{n^2}{\bar{l}}\leq\frac{abn}{\sqrt{1-a^2b^2}}$ two cases may arise, $r_0^-<r_-<0<M$ and $r_0^-<0<r_-<M$.
In both scenarios the curvature singularity is covered by the event horizon $r_-$, then a black hole with one event horizon locally isomorphic to $\mathcal{H}^2$ is identified. \\
\\
Case II: The pole of the conformal factor does not belong to the radial coordinate domain $(r_0^-,\infty)$, $r=-|M|<r_0^-=-|M|+|ab|\sqrt{n^2+|M|^2}$. $r_0^-$ is always greater than $r_-$ and $r_{--}$. Then, for values of the mass parameter $\frac{|ab|n}{\sqrt{1-|ab|^2}}\leq\frac{n^2}{\bar{l}}-\frac{\bar{l}}{4}<|M|$ the two Killing horizons $r_+$ and $r_{++}$ emerge defining three cases: (i) $r_0^-<r_+<0<r_{++}$, (ii) $r_0^-<0<r_+<r_{++}$, and (iii) $r_0^-<r_{++}$. The first two cases define a stationary black hole with inner and event horizons while the third case represents a single event horizon black hole.\\
\\
Case III: Here we have $r=-|M|>r_0^-=-|M|-ab\sqrt{n^2+M^2}$ and thus the spacetime divides into the regions $\mathcal{R}^{-}\in(r^{-}_0,-|M|)$ and $\mathcal{R}^{+}\in(-|M|,\infty)$.
Therefore:
\vspace{0.2cm}
\\
$\mathcal{R}^{+}$: This region is bounded by $r=-|M|$ and $r=\infty$. For $|M|>\frac{n^2}{\bar{l}}-\frac{\bar{l}}{4}$ the following two cases display:  (i) $-|M|<0<r_+<r_{++}$ and (ii) $-|M|<r_+<0<r_{++}$, both representing stationary regular black holes with a similar causal structure to the one of the Reissner-Nordstr\"om anti--de Sitter black hole, see Fig. 7. By saturating the mass bound both horizons merge and the solution represents a regular black hole with one event horizon, Fig. 9. Finally, if  $0<|M|<\frac{n^2}{\bar{l}}-\frac{\bar{l}}{4}$ the spacetime is devoid of Killing horizons and it transforms into a wormhole connecting two anti--de Sitter boundaries. See Fig. 12. The following anti--de Sitter black bounce is thus observed
\begin{equation} AdS\hspace{0.1cm}black\hspace{0.1cm}bounce \rightarrow
\left\{ \begin{aligned} 
\frac{n^2}{\bar{l}}-\frac{\bar{l}}{4}<&|M|<\infty\hspace{1.45cm} \rightarrow \hspace{0.3cm} regular\hspace{0.1cm}black\hspace{0.1cm}hole\hspace{0.1cm}with\hspace{0.1cm}inner\hspace{0.1cm}and\hspace{0.1cm}event\hspace{0.1cm}horizons\\
&|M|=\frac{n^2}{\bar{l}}-\frac{\bar{l}}{4}\hspace{0.66cm} \rightarrow  \hspace{0.3cm}regular\hspace{0.1cm}black\hspace{0.1cm}hole\hspace{0.1cm}with\hspace{0.1cm}one\hspace{0.1cm}event\hspace{0.1cm}horizon\\
0<&|M|<\frac{n^2}{\bar{l}}-\frac{\bar{l}}{4}\hspace{0.66cm} \rightarrow  \hspace{0.3cm}wormhole.
\end{aligned} \right. 
\end{equation}
\begin{figure}[h!]
    \includegraphics[width=1.01\textwidth]{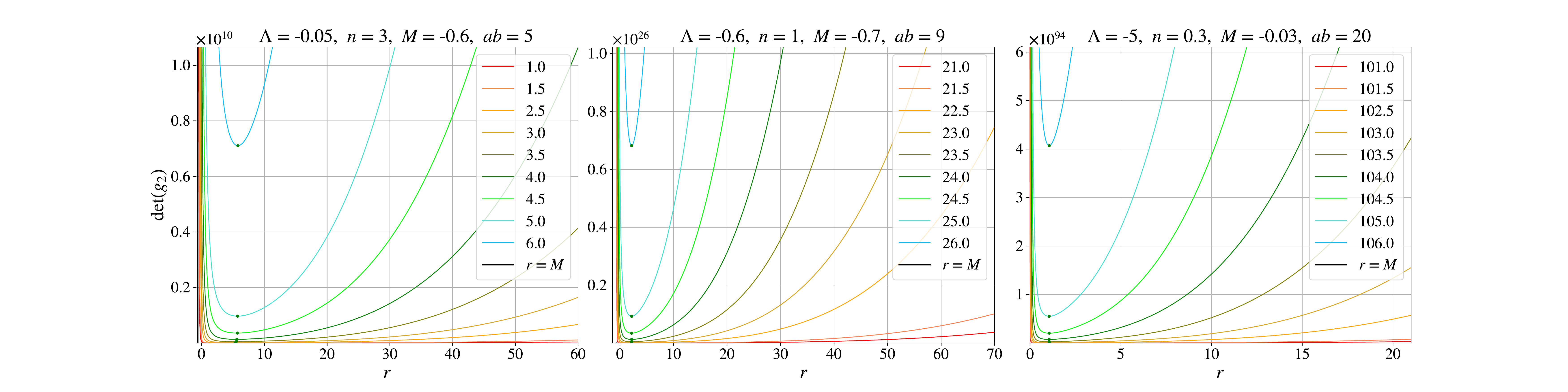}
    \caption{Position of the wormhole throats for different set of parameters $\Lambda$, $M$, and $n$ and for several values of the coordinate $\theta$.}
    \label{fig12}
\end{figure}
\\
$\mathcal{R}^{-}$: If the mass parameter satisfies $|M|<\frac{\bar{l}}{4}-\frac{n^2}{\bar{l}}$ then both horizons $r_-$ and $r_{--}$ appear. The following two cases are displayed: (i) $r_0^-<r_{--}<r_-<-|M|$ and (ii) $r_0^-<r_-<-|M|$. Both cases represent black holes, with inner and event or only event horizons, respectively.

\subsubsection{$\Lambda>0$ and $K=-1$}

The combination $\Lambda>0$ and $K=-1$ always provides a complex $\bar{l}$. Therefore, the spacetime is devoid of Killing horizons and the curvature singularity as well as the pole of the conformal factor are, in principle, naked. This situation will occur for all the cases contained in (\ref{poscuv}) and (\ref{negcuv}) in which the pole of the conformal factor does not belong to the corresponding domains of the radial coordinate. On the other hand, for all cases in which the spacetime is divided into the regions $\mathcal{R}^{-}$ and $\mathcal{R}^{+}$ the situation differs. As a matter of fact, the curvature singularity and the pole of the scalar field will always belong to the region $\mathcal{R}^{-}$ and in consequence region $\mathcal{R}^{+}$ will be free of pathologies. There is only one possible configuration, a bouncing cosmology connecting two de Sitter Universes with cosmological constants $\frac{\Lambda}{3a^2b^2}$ and $\frac{\Lambda}{3}$, no matter the value of the mass parameter.
This can be directly observed from the determinant of the two dimensional $t=r=constant$ sections (\ref{dethip}). The metric function is everywhere negative and in consequence this determinant never vanishes. This will be observed in three of the cases previously studied, Case I when $r_0^+$ takes places and Cases I and III when $r_0^-$ appears.
Here we provide a graphical proof of the bounce for the case in which $r_0^+$ takes place. See Fig. 13. Due to the hyperbolic nature of the transverse section this geometry is free of Misner strings, and moreover does not possess any kind of closed timelike curves as long as $\Lambda>3/4n^2$, see the seed analysis. 
\begin{figure}[h!]
    \includegraphics[width=1.01\textwidth]{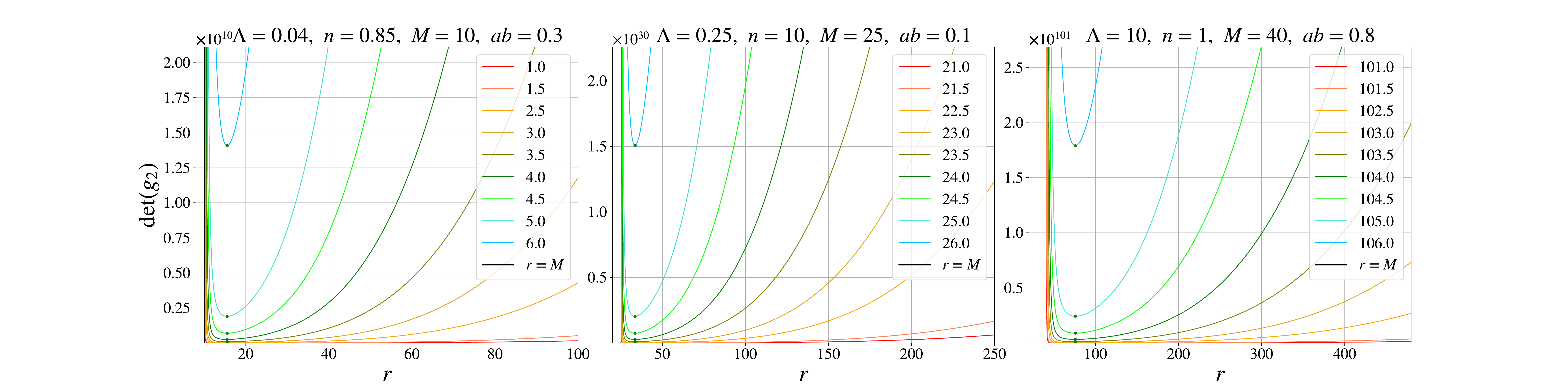}
    \caption{Position of the bounce for different set of parameters $\Lambda$, $M$ and $n$ and for several values of the coordinate $\theta$.}
    \label{fig13}
\end{figure}
\\
\subsubsection{$\Lambda<0$ and $K=1$}

This combination possesses complex $\bar{l}$ as well as in the previous case. No Killing horizons take place and the curvature singularity and scalar field pole will be naked in all those cases in which the conformal factor pole does not belong to the domain of the radial coordinate. For those cases in which the spacetime is divided into the regions $\mathcal{R}^{-}$ and $\mathcal{R}^{+}$ the spacetime $\mathcal{R}^+$ will represent an anti--de Sitter wormhole connecting to asymptotic boundaries with constant curvatures $\frac{\Lambda}{3a^2b^2}$ and $\frac{\Lambda}{3}$. For a non geodesic observer the Misner string is unavoidable as well as the occurrence of closed timelike curves. As a consequence of the existence of closed timelike curves the throat of the wormhole can be graphically observed only for a narrow set of parameters.
See Fig. 14 for Case I with $r_0^+$.
\begin{figure}[h!]
    \includegraphics[width=1.01\textwidth]{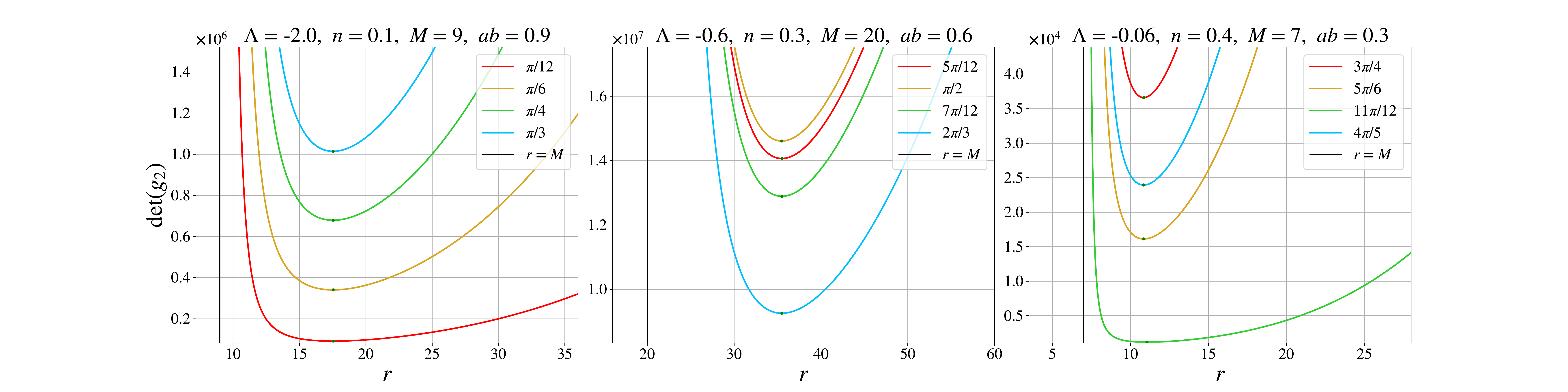}
    \caption{Position of the wormhole throat for different set of parameters and for different values of the polar coordinate. Notice that for this case it is not possible to access the wormhole throat for $\theta=\pi$.}
    \label{fig14}
\end{figure}

\subsubsection{$K=0$ \rm{cases}}

For the cases in which the base manifold is flat the absence of Killing horizons always takes place. Then, the metric function is either everywhere negative or everywhere positive depending if the cosmological constant is positive or negative. Again, generically the curvature singularity and the scalar field pole might be naked unless the conformal factor includes a new asymptotic region that provides a spacetime $\mathcal{R}^{+}$ for which no singularities appear. In such cases the solution is interpreted as a bouncing cosmology connecting to de Sitter Universes or as a wormhole connecting two locally anti--de Sitter boundaries. The Misner string is not present, for planar base manifold geometries these are pulled out to infinity. The presence of closed timelike curves is a generic feature for nongeodesic observers. These are not going to affect the cosmological solution, nevertheless severally restricts the existence of the throat in the wormhole case, narrowing the set of parameters in which the latter might take place.

\section{Conclusions and further developments} 

In the present paper we have constructed and analyzed a new family of Taub-NUT spacetimes representing regular black holes, wormholes, and bouncing cosmologies in Einstein-$\Lambda$ theory in which a self-interacting conformally coupled scalar field is considered. The self-interaction deviates from the usual conformal potential by including all power-counting super-renormalizable contributions. The construction is based on the use of a specific set of transformations \cite{Ayon-Beato:2015ada}, that once applied to a seed solution that solves the field equations of a conformally coupled scalar-tensor theory with a conformal potential (\ref{seedEOM}), provides a new set of geometries that drastically enlarge the causal structure of the spacetime. These solutions solves the new set of field equations (\ref{transEOM}) where the scalar field action enjoys conformal invariance up to the presence of a nonconformal potential that exhibits all power-counting superrenormalizable contributions. Additionally, the cosmological constant is also modified. Comparison with the seed spacetime is direct. Table I summarizes our most relevant findings.\\
\\
The first four solutions contained in Table I possess a spherical foliation and the seed cosmological constant is positive. In this scenario the seed solution represents a Taub-NUT black hole in de Sitter space with three Killing horizons. Its black hole interpretation is subjected to the arguments of \cite{Clement:2015cxa,Clement:2015aka}, this due to the presence of the Misner string and closed timelike curves which might generically appear. Then, we adopted this interpretation for geodesic observers only. Our new solutions drastically differ from this causal structure.
The presence of the conformal factor pole divides the spacetime into two regions from which the most relevant is $\mathcal{R}^+$. This region is interpreted as a regular black hole with two horizons only, the event horizon $r_+$ and the cosmological horizon $r_{++}$, which by smoothly moving the mass parameter transits to a bouncing cosmology connecting two de Sitter Universes. It is shown the existence of an intermediate phase in which a de Sitter wormhole or a de Sitter bouncing cosmology might appear, this under the presence of a cosmological horizon. Less attractive is region $\mathcal{R}^-$, which mostly represents naked singularities with the exception of Case III in the presence of $r_0^-$. This case represents a de Sitter black hole with event $r_+$ and cosmological $r_{++}$ horizons. 
It is important to stress that for all the cosmological configurations the absence of closed timelike curves is ensured beyond geodesic observers. Nonetheless, the Misner string is still visible for them. All these solutions are completely regular in terms of their curvature invariants and the scalar field profile is everywhere nondivergent. \\
\\
The next six solutions contained in Table I possess a hyperbolic foliation and the seed cosmological constant is negative. Among all the solutions contained in the seed metric this is the most relevant. For hyperbolic foliation this spacetime does not contain a Misner string, and moreover it is shown that by properly restricting the cosmological constant in terms of the NUT parameter it is possible to remove any trace of closed timelike curves. This geometry represents an anti--de Sitter regular black hole with three horizons. Our solutions enhance this geometry in the following manner: again, the presence of a pole in the conformal factor implies the division of the spacetime into the regions  $\mathcal{R}^-$ and $\mathcal{R}^+$. Region $\mathcal{R}^+$ represents a regular black hole with inner and event horizons $r_+$ and $r_{++}$, which by smoothly moving the mass parameter transits to a wormhole configuration that connects two anti--de Sitter asymptotic boundaries with different constant curvatures. An intermediate phase given by a regular black hole with a single event horizon $r_H$ is observed. This phase might appear in two different setups, either as a regular black hole inside an anti--de Sitter wormhole or as a regular black hole of which its interior is given by a cosmological bounce. In both scenarios this phase shows an interesting characteristic: it possesses only one horizon, the event horizon, and in consequence is free of Cauchy horizons which are known for suffering from instabilities. 
On the other hand, region $\mathcal{R}^-$ provides anti--de Sitter black holes with inner and event horizons. It is important to stress that all these solutions are free of Misner strings and of any closed timelike curves for any observer, geodesic or nongeodesic. \\
\\
Other solutions might be highlighted as well, nonetheless none of them representing black bounce transitions. These are the case with positive cosmological constant and hyperbolic base manifold that represents a bouncing cosmology and the case with negative cosmological constant and spherical base manifold that represents a wormhole. The latter case being valid for a restricted set of the parameter space. Additionally, several cases represent black holes with different horizons distributions when no pole of the conformal factor takes place. These latter are not included in Table I. \\
\\
An important aspect to be discussed in all these solutions is whether or not they satisfy reasonable energy conditions. Due to the complexity of our solutions  explicit expressions for the energy density or for the pressures are rather involved, nevertheless some general conclusions can be drawn by casting the energy momentum tensor of our matter fields in the form of a perfect fluid, in this case an anisotropic fluid which in an orthonormal frame is given by $T^{ab}:=(\rho, p_1,p_2,p_2)$. It is realized to be useful to consider the case in which the electric charge vanishes. From the point of view of the solutions this does not produce any substantial change, besides a fine tune between the cosmological constant $\Lambda$ and the conformal coupling $\alpha$. 
It is observed that the null energy condition given by $\rho+p_1\geq0$ and $\rho+p_2\geq0$ is always satisfy for bouncing cosmologies. On the other hand, wormhole configurations requires, as usual, the presence of exotic matter, which in our cases is represented by our scalar fields not obeying the aforementioned energy condition. For black holes energy conditions are not satisfied everywhere due to the change of sign of the metric function $F(r)$. However, several cases respect the null energy condition in the domain of outer communications.\\ 
\\
Many appealing directions might be further explored. Remaining in the context of Taub-NUT spacetimes it would be interesting to study the Euclidean version of our solutions. These gravitational instantons will certainly exhibit interesting new geometric and topological features healing the pathologies associated with the Misner string and the closed timelike curves \cite{eu-case}. Additionally, either in their Lorentzian or Euclidean versions, our solutions offer an interesting arena to study black hole thermodynamic \cite{Mann:1999pc,Garfinkle:2000ms,Ciambelli:2020qny,Hennigar:2019ive,Durka:2019ajz,Frodden:2021ces}. It would be worth studying the existence of phase transitions in the presence of scalar fields not only in standard black hole thermodynamic but also in the context of black hole chemistry \cite{Simovic:2020dke,Johnson:2014pwa} and to understand how the presence or absence of the Misner string affects the usual charges.
On the other hand it is clear that the set of transformations defined in \cite{Ayon-Beato:2015ada} do not depend on the particular symmetries the metric or the scalar field enjoy. Namely, for any solution of a conformally coupled theory with a conformal potential a new conformally related family of the type found here can be constructed. A concrete example is to take the C-metric solution with a conformally coupled scalar found in \cite{Charmousis:2009cm} and construct a new family of accelerating geometries. The transformations affect in a highly nontrivial manner the geometric structure of this spacetime providing interesting new phenomena to be explored \cite{dolphiprogress}. Even more, a full new Pleba\'nski-Demia\'nski family of solutions can be constructed along the lines of \cite{Anabalon:2009qt} to explore how the transformation modifies the corresponding limits that provide specific spacetimes with a desirable specific symmetry. It is known that the Pleba\'nski-Demia\'nski solutions found in \cite{Anabalon:2009qt} does not provide a healthy limit when turning off the acceleration while keeping rotation, this in the presence of the cosmological constant. An interesting path to follow would be to investigate whether or not this limit might be cured by the effect of the transformations. 
This can be pushed further by including higher order corrections to the original action that do not spoil the initial conformal symmetry of the seed theory \cite{Cisterna:2021xxq}. The same conclusions are valid for solutions with external magnetic fields \cite{Astorino:2013xc,Astorino:2013sfa}, for pp/AdS-waves metrics \cite{Ayon-Beato:2005gdo} or conformally coupled solutions in higher or lower dimensions \cite{Oliva:2011np,Giribet:2014bva,Galante:2015voa}. \\
\\
Finally, specifically in the context of the black bounce solutions described here it would be interesting to explore all kinds of applications. The most obvious, again, black hole thermodynamic since now they represent solutions with a sensible action principle with a clear field content\footnote{Recently, following a bottom up approach these type of models have been embedded in higher order scalar-tensor theories \cite{Chatzifotis:2021hpg}. In there, the parameter that controls the bounce emerges in the action principle when integrating out the theory that supports the desired solution. Similar strategy has been also applied in the context of nonlinear electrodynamics \cite{Canate:2022gpy}.}. On the other hand, it is possible to study several astrophysically relevant aspects of these black holes, as for example black hole shadows \cite{Cunha:2017eoe,Lima:2021las}, gravitational lensing \cite{Virbhadra:1999nm,Tsukamoto:2020bjm}, black hole mimickers \cite{Johnson-Mcdaniel:2018cdu,Mazza:2021rgq}, quasinormal modes and echoes \cite{Churilova:2019cyt,Yang:2021cvh,Wu:2022eiv,Cano:2021qzp}, to mention a few examples. 
We expect to report along these lines in our upcoming contributions. 

\section{Acknowledgments} 
We appreciate interesting discussions with Marco Astorino, Crist\'obal Corral, Gaston Giribet, Julio Oliva, Julio Méndez-Zavaleta and Marcello Ortaggio. A.C would like to acknowledge the Institute of Mathematics of the Czech Academy of Science for its kind hospitality during the development of this project. A.C work is funded by FONDECYT Regular grants No. 1210500, No. 1181047 and Beca Chile de Postdoctorado Grant No. 74200012. J.B is supported by the  ``Programme to support prospective
human resources – post Ph.D. candidates''  of the Czech Academy of Sciences, project No. L100192101. N.M is funded by Beca ANID de Doctorado Nacional 2021 Grant No. 21212393. A.V is supported in part by by MIUR-PRIN contract No. 2017CC72MK003.

\small
\begin{table}[h!]
\centering
\begin{tabular}{|m{3.5cm} | l | l| m{6.7cm} |}
\hline 
$(K,\Lambda)$   & Singularity &Region/Mass & Interpretation\\
\hline 
\hline 

$\begin{matrix}
K=1\\
0<\Lambda < \dfrac{3}{8n^2}
\end{matrix}$  &  $r_0^+$ Case I  & $\mathcal{R}^+:\left\{\begin{matrix}
\dfrac{abn}{\sqrt{1-a^2b^2}}<M<\dfrac{\bar{l}}{4}-\dfrac{n^2}{\bar{l}}\\
M=\dfrac{\bar{l}}{4}-\dfrac{n^2}{\bar{l}}
\\
\dfrac{\bar{l}}{4}-\dfrac{n^2}{\bar{l}}<M<\infty
\end{matrix}  \right.
$   &  \textit{de Sitter black bounce: regular black hole with event $r_+$ and cosmological $r_{++}$ horizons, de Sitter wormhole or bouncing cosmology with a single cosmological horizon $r_c$ and free horizons bouncing cosmology.}\\
\hline
$\begin{matrix}
K=1\\
0<\Lambda < \dfrac{3}{8n^2}
\end{matrix}$  &  $r_0^-$ Case I  & $\mathcal{R}^+:\left\{\begin{matrix}
0<M<\dfrac{\bar{l}}{4}-\dfrac{n^2}{\bar{l}}\leq \dfrac{abn}{\sqrt{1-a^2b^2}}\\
M=\dfrac{\bar{l}}{4}-\dfrac{n^2}{\bar{l}}
\\
\dfrac{\bar{l}}{4}-\dfrac{n^2}{\bar{l}}<M<\dfrac{abn}{\sqrt{1-a^2b^2}}
\end{matrix}  \right.
$   &  \textit{de Sitter black bounce: regular black hole with event $r_+$ and cosmological $r_{++}$ horizons, de Sitter wormhole or bouncing cosmology with a single cosmological horizon $r_c$ and free horizons bouncing cosmology.}\\
\hline
$\begin{matrix}
K=1\\
\dfrac{3}{8n^2}<\Lambda < \dfrac{3}{4n^2}
\end{matrix}$  &  $r_0^-$ Case III  & $\mathcal{R}^+:\left\{\begin{matrix}
\dfrac{n^2}{\bar{l}}-\dfrac{\bar{l}}{4}<\left|M\right|<\infty\\
\left|M\right|=\dfrac{n^2}{\bar{l}}-\dfrac{\bar{l}}{4}
\\
0<\left|M\right|<\dfrac{n^2}{\bar{l}}-\dfrac{\bar{l}}{4}
\end{matrix}  \right.
$   &  \textit{de Sitter black bounce : regular black hole with event $r_+$ and cosmological $r_{++}$ horizons, de Sitter wormhole or bouncing cosmology with a single cosmological horizon $r_c$ and free horizons bouncing cosmology.}\\

\hline
$\begin{matrix}
K=1\\
0<\left|\Lambda\right| <\dfrac{3}{8n^2} 
\end{matrix}$  &  $r_0^-$ Case III  & $\mathcal{R}^-:\left\{\left|M\right|<\dfrac{\bar{l}}{4}-\dfrac{n^2}{\bar{l}}  \right.
$   &  \textit{de Sitter black hole with two horizons, event $r_+$ and cosmological $r_{++}.$
}\\

\hline

$\begin{matrix}
K=-1\\
0<\left|\Lambda\right| < \dfrac{3}{8n^2}
\end{matrix}$  &  $r_0^+$ Case I  & $\mathcal{R}^+:\left\{\begin{matrix}
0<M<\dfrac{\bar{l}}{4}-\dfrac{n^2}{\bar{l}}\\
M=\dfrac{\bar{l}}{4}-\dfrac{n^2}{\bar{l}}
\\
M>\dfrac{\bar{l}}{4}-\dfrac{n^2}{\bar{l}}
\end{matrix}  \right.
$   &  \textit{anti--de Sitter black bounce: regular black hole with inner $r_+$ and event $r_{++}$ horizons,
regular black hole with one event horizon $r_H$ either inside of a wormhole or with an interior cosmological bounce, and anti--de Sitter wormhole.}\\
\hline
$\begin{matrix}
K=-1\\
0<\left|\Lambda\right| < \dfrac{3}{8n^2}
\end{matrix}$  &  $r_0^-$ Case I  & $\mathcal{R}^+:\left\{\begin{matrix}
0<M<\dfrac{\bar{l}}{4}-\dfrac{n^2}{\bar{l}}\leq\dfrac{abn}{\sqrt{1-a^2b^2}}\\
M=\dfrac{\bar{l}}{4}-\dfrac{n^2}{\bar{l}}
\\
\dfrac{\bar{l}}{4}-\dfrac{n^2}{\bar{l}}<M<\dfrac{abn}{\sqrt{1-a^2b^2}}
\end{matrix}  \right.
$   &  \textit{anti--de Sitter black bounce: regular black hole with inner $r_+$ and event $r_{++}$ horizons,
regular black hole with one event horizon $r_H$ either inside of a wormhole or with an interior cosmological bounce, and anti--de Sitter wormhole.}\\
\hline
$\begin{matrix}
K=-1\\
\dfrac{3}{8n^2}<\left|\Lambda\right| <\dfrac{3}{4n^2} 
\end{matrix}$  &  $r_0^-$ Case III  & $\mathcal{R}^+:\left\{\begin{matrix}
\dfrac{n^2}{\bar{l}}-\dfrac{\bar{l}}{4}<\left|M\right|<\infty\\
\left|M\right|=\dfrac{n^2}{\bar{l}}-\dfrac{\bar{l}}{4}
\\
0<\left|M\right|<\dfrac{n^2}{\bar{l}}-\dfrac{\bar{l}}{4}
\end{matrix}  \right.
$   &  \textit{anti--de Sitter black bounce: regular black hole with inner $r_+$ and event $r_{++}$ horizons,
regular black hole with one event horizon $r_H$ either inside of a wormhole or with an interior cosmological bounce, and anti--de Sitter wormhole.
}\\

\hline
$\begin{matrix}
K=-1\\
0<\left|\Lambda\right| < \dfrac{3}{8n^2}
\end{matrix}$  &  $r_0^+$ Case I  & $\mathcal{R}^-:\left\{0<M<\dfrac{\bar{l}}{4}-\dfrac{n^2}{\bar{l}}\right.
$   &  \textit{anti--de Sitter black hole with a single event horizon locally isomorphic to $\mathcal{H}^2$: $r^{-}0 < r- < 0 < M$.}\\
\hline
$\begin{matrix}
K=-1\\
0<\left|\Lambda\right| < \dfrac{3}{8n^2}
\end{matrix}$  &  $r_0^-$ Case I  & $\mathcal{R}^-:\left\{0<M<\dfrac{\bar{l}}{4}-\dfrac{n^2}{\bar{l}}\leq \dfrac{abn}{\sqrt{1-a^2b^2}}\right.
$   &  \textit{anti--de Sitter black hole with a single event horizon locally isomorphic to $\mathcal{H}^2$. Two cases: $r^{-}0 < r- < 0 < M$ and  $r^{-}0 < 0 <  r- < M$.}\\
\hline
$\begin{matrix}
K=-1\\
0<\left|\Lambda\right| < \dfrac{3}{8n^2}
\end{matrix}$  &  $r_0^-$ Case III  & $\mathcal{R}^-:\left\{\left|M\right|<\dfrac{\bar{l}}{4}-\dfrac{n^2}{\bar{l}}\right.
$  &  \textit{anti--de Sitter black hole with one or two horizons locally isomorphics to $\mathcal{H}^2$. Two cases: $r^{-}0 < r{--} < r_- < -\left|M\right|$ and  $r^{-}0 <  r- < -\left|M\right|$.}\\
\hline
$\begin{matrix}
 K=-1
, \Lambda > \dfrac{3}{4n^2}
\end{matrix}$  &  $r_0^+$ Case I  & $\mathcal{R}^+:\left\{M>0\right.
$  &  \textit{de Sitter bouncing cosmology}\\
\hline
$\begin{matrix}
 K=1
, \Lambda <0
\end{matrix}$  &  $r_0^+$ Case I  & $\mathcal{R}^-:\left\{M>0\right.
$  &  \textit{anti--de Sitter wormhole.}\\
\hline

    \end{tabular}
\caption{Summary of the most significant results for different values of $K$ and $\Lambda$.}
\end{table}

\section{Appendix: $\Lambda=0$ case}

When the seed cosmological constant vanishes consistence of the seed solution (\ref{charmetric}) requires the absence of the seed quartic self-interaction, $\alpha=0$. The seed scalar field integrates in a slightly different manner. In this case no potential $V(\bar{\phi})$ appears after the transformation neither an effective cosmological constant $\lambda$. The lapse function takes a very simple form possessing one single root at $r=M$, same point at which the conformal factor has a pole. The solution reads 
\begin{equation}
\begin{aligned}
d\bar{s}^2&=\frac{(a\sqrt{n^2+M^2-\frac{\kappa \bar{Q}^2}{2(1-a^2)}}+r-M)^2}{(r-M)^2}\left[-\frac{K(r-M)^2}{r^2+n^2}(dt+\mathcal{B})^2+\frac{dr^2}{\frac{K(r-M)^2}{r^2+n^2}}+(r^2+n^2)d\Sigma_{K}^2\right], \hspace{0.1cm} \bar{A}=\frac{\bar{Q}r}{r^2+n^2}(dt+\mathcal{B})\\
\\
\bar{\phi}(r)&=\left(\frac{6}{\kappa}\right)^{1/2}\frac{\sqrt{n^2+M^2-\frac{\kappa \bar{Q}^2}{2(1-a^2)}}+a(r-M)}{a\sqrt{n^2+M^2-\frac{\kappa\bar{Q}^2}{2(1-a^2)}}+r-M}. 
\end{aligned} 
\end{equation}
As explained before the zero of the conformal factor induces a curvature singularity, now located at
\begin{equation}
r_0=M-a\sqrt{n^2+M^2-\frac{\kappa\bar{Q}^2}{2(1-a^2)}},  \label{singuflat} 
\end{equation}
which, depending on the signs of $M$ and $a$ will be either located in the positive or negative range of the radial coordinate. The electric charge must accomplish $\bar{Q}^2<\frac{2(1-a^2)(n^2+M^2)}{\kappa}$ such that the conformal factor remains real. 
In comparison with the previous cases in which the cosmological constant was nonvanishing here the presence of the conformal factor pole is fundamental, otherwise the spacetime will always exhibit a naked singularity. The region $\mathcal{R}^+$ will represent either a wormhole or a cosmological bounce depending on the horizon geometry. No black hole horizons will exists. This solution represents the charged Taub-NUT extension of Barcelo's wormhole \cite{Barcelo:2000zf} for $K=1$. Region $\mathcal{R}^{-}$ contains a naked singularity.

\end{document}